\documentclass[fleqn,usenatbib]{mnras}
\usepackage[T1]{fontenc}

\DeclareRobustCommand{\VAN}[3]{#2}
\let\VANthebibliography\thebibliography
\def\thebibliography{\DeclareRobustCommand{\VAN}[3]{##3}\VANthebibliography}

\usepackage{graphicx}	% Including figure files
\usepackage{amsmath}	% Advanced maths commands
\usepackage{amssymb}
\usepackage{ulem}
\usepackage{tikz}
\usepackage{newtxtext,newtxmath}

\newcommand{\jetcaf}{{\fontfamily{qcr}\selectfont JeTCAF}}
\newcommand{\tcaf}{{\fontfamily{qcr}\selectfont TCAF}}

\newcommand{\tbabs}{{\fontfamily{qcr}\selectfont TBABS}}
\newcommand{\pcfabs}{{\fontfamily{qcr}\selectfont PCFABS}}
\newcommand{\zxipcf}{{\fontfamily{qcr}\selectfont ZXIPCF}}

\newcommand{\diskbb}{{\fontfamily{qcr}\selectfont DISKBB}}

\title[Spectral and Temporal Studies of Swift\,J1658.2--4242]{Spectral and Temporal Studies of Swift\,J1658.2--4242 using {\it AstroSat} Observations with {\tt JeTCAF} Model}

\author[S. Mondal and V. Jithesh]{
Santanu Mondal$^{1}$\thanks{E-mail: santanuicsp@gmail.com (SM)} and
V. Jithesh$^{2}$\thanks{E-mail: jithesh.v@christuniversity.in (VJ)}\thanks{Equal First Author}
\\
$^{1}$Indian Institute of Astrophysics, II Block Koramangala, Bengaluru 560034, India \\
$^{2}$Department of Physics and Electronics, CHRIST (Deemed to be University), Hosur Main Road, Bengaluru - 560029, India \\
}

\date{Accepted XXX. Received YYY; in original form ZZZ}

\pubyear{2023}

\begin{document}
\label{firstpage}
\pagerange{\pageref{firstpage}--\pageref{lastpage}}
\maketitle

\begin{abstract}
We present the X-ray spectral and temporal analysis of the black hole X-ray transient Swift\,J1658.2--4242 observed by {\it AstroSat}. Three epochs of data have been analysed using the \jetcaf\,model to estimate the mass accretion rates and to understand the geometry of the flow. The best-fit disc mass accretion rate ($\dot m_d$) varies between $0.90^{+0.02}_{-0.01}$ to $1.09^{+0.04}_{-0.03}$ $\dot M_{\rm Edd}$ in these observations, while the halo mass accretion rate changes from $0.15^{+0.01}_{-0.01}$ to $0.25^{+0.02}_{-0.01}$ $\dot M_{\rm Edd}$. We estimate the size of the dynamic corona, that varies substantially from $64.9^{+3.9}_{-3.1}$ to $34.5^{+2.0}_{-1.5}$ $r_g$ \textcolor{black}{and a moderately high  jet/outflow collimation factor stipulates isotropic outflow.} The inferred high disc mass accretion rate and bigger corona size indicate that the source might be in the  intermediate to soft spectral state of black hole X-ray binaries. The mass of the black hole estimated from different model combinations is $\sim 14 M_\odot$. In addition, we compute the quasi-periodic oscillation (QPO) frequencies from the model-fitted parameters, which match the observed QPOs. We further calculate the binary parameters of the system from the decay profile of the light curve and the spectral parameters. The estimated orbital period of the system is $4.0\pm0.4$ hr by assuming the companion as a mid or late K-type star. Our analysis using the \jetcaf\, model sheds light on the physical origin of the spectro-temporal behaviour of the source, and the observed properties are mainly due to the change in both the mass accretion rates \textcolor{black}{and absorbing column density}.

\end{abstract}


%%%%%%%%%%%%%%%%%%%%%%%%%%%%%%%%%%%%%%%%%%%%%%%%%%

%%%%%%%%%%%%%%%%% BODY OF PAPER %%%%%%%%%%%%%%%%%%

\section{Introduction} \label{sec:intro}

Accretion onto black hole (BH) in the X-ray binaries produces electromagnetic radiation at different energy bands. In the X-ray band, almost all black hole X-ray binaries (BHXRBs) show spectral variability with different spectral states from hard (HS) to soft (SS) through intermediate states \citep[][and references therein]{HomanEtal2001ApJS..132..377H,BelloniEtal2005A&A...440..207B,RemillardMcClin2006ARA&A..44...49R,NandiEtal2012A&A...542A..56N,DebnathElal2015MNRAS.447.1984D,ShuiEtal2021MNRAS.508..287S,BabyEtal2021MNRAS.508.2447B}. These spectral states change can occur due to several reasons, e.g., change in mass accretion rate or corona \citep{ChakTitarchuk1995ApJ...455..623C,NarayanEtal1996ApJ...457..821N,LaurentTitarchuk2001ApJ...562L..67L,RemillardMcClin2006ARA&A..44...49R,MondalEtal2014ApJ...786....4M,ChatterjeeEtal2016ApJ...827...88C}, accretion disc instabilities, and/or hysteresis behaviour of the cooling and viscosity \citep{JaniukEtal2002ApJ...576..908J,Meyer-HofmeisterEtal2005A&A...432..181M,MondalEtal2017ApJ...850...47M} during the outburst phase. In the long term outbursts or between outbursts, flaring and re-flaring events are observed, which can also be explained from the change in viscosity of accreting matter \citep{ChakrabartiEtal2019AdSpR..63.3749C}.

In the temporal domain too, the sources show variability mainly in the form of different types of quasi-periodic oscillations (QPOs), whose frequency ranges from mHz (low frequency; LF) to kHz (high frequency; HF). The LFQPOs can be divided into three types (type A, B, and C) depending on their root mean square (rms) and Q values \citep{CasellaEtal2005ApJ...629..403C}. Type C QPOs are mainly identified in the hard-intermediate state (HIMS), while type A and B QPOs are detected in the soft-intermediate state \citep[SIMS;][]{WijnandsEtal1999ApJ...526L..33W, CasellaEtal2005ApJ...629..403C, MottaEtal2011MNRAS.418.2292M}. Several models have been proposed to explain the origin and evolution of LFQPOs in X-ray binaries. The proposed models are generally based on three different mechanisms: oscillation of the shock or dynamic corona \citep{MolteniEtal1996ApJ...457..805M, ChakrabartiEtal2015MNRAS.452.3451C}, Lense-Thirring precession \citep{StellaVietri1998ApJ...492L..59S, StellaEtal1999ApJ...524L..63S,IngramEtal2009MNRAS.397L.101I}, and instabilities \citep[e.g.][]{TaggerPellat1999A&A...349.1003T, TitarchukFiorito2004ApJ...612..988T, CabanacEtal2010MNRAS.404..738C}. LFQPOs are often detected in transient BHXRBs and their origin provides essential information regarding the accretion properties and accretion flow geometry around the BH. The appearance and disappearance of these QPOs are also interlinked with the spectral states, indicating that both spectral and temporal behaviours originated from the same system and geometrical flow configuration. Therefore, it is worth exploring the models which can explain both spectral and timing signatures simultaneously.

{\it AstroSat} \citep{SinghEtal2014SPIE.9144E..1SS, AgrawalEtal2017JApA...38...30A} is India’s first multi-wavelength astronomical observatory, which contains five instruments onboard: Soft X-ray Telescope (SXT), Large Area X$-$ray Proportional Counter (LAXPC), Cadmium Zinc Telluride Imager (CZTI), Scanning Sky Monitor (SSM) and an Ultra-Violet Imaging Telescope (UVIT). Among them, LAXPC is ideal for the rapid time variability studies of X-ray binaries with moderate spectral capabilities. The high sensitivity and medium resolution spectral capability of SXT in the 0.3--8 keV energy band are useful for broadband spectral studies simultaneously with LAXPC. These capabilities provided a unique opportunity to investigate the spectro-timing properties of several X-ray binaries (including new transients), particularly BHXRBs using {\it AstroSat}. The analysis of the first {\it AstroSat} data sets of Cygnus X--1 identified the source in the canonical hard state characterised by a prominent thermal Comptonisation component having a photon index of $\sim 1.8$ and the coronal temperature ($\rm kT_{e}$) of $> 60$ keV along with signatures of weak reflection. The temporal analysis revealed two broad peaks in the power density spectrum. For both components, the fractional rms exhibited a decreasing trend with energy, while the time lag showed an increasing trend with energy \citep{Mis17}. A more detailed study with six {\it AstroSat} observations, \citet{Maq19} estimated the energy-dependent fractional rms and time lag at different frequencies. In addition, by invoking a fluctuation propagation model with simple truncated disk geometry, they explained the mechanism responsible for the observed timing properties of the source. {\it AstroSat} observations of MAXI J1535$-$571 identified the source in the HIMS of BHXRBs. Using the broadband spectrum with reflection models, \citet{Sri19} estimated the spin, BH mass and distance to the source, and the values are 0.67, 10.4 $M_\odot$ and 5.4 kpc, respectively. From the timing analysis, \citet{Bha19} reported the presence of strong QPOs, which vary in the range of 1.7--3.0 Hz. A tight correlation between the QPO frequency and the power-law spectral index has been observed, while the QPO frequency is not correlated with the flux. From the strong correlation, the authors argue that the QPOs are closely related to the Comptonising region rather than the accretion disk. 

The independent observations of MAXI J1820+070 with {\it AstroSat} and contemporaneous {\it NuSTAR} revealed an inhomogeneous corona characterised by high coronal temperature \citep{Cha20}. Moreover, the broadband coverage using the three simultaneously observing X-ray instruments (SXT, LAXPC and CZTI) onboard {\it AstroSat} helps to estimate the BH mass of the source, which is in the range of 6.7--13.9 $M_\odot$. {\it AstroSat} observed GRS 1915+105 multiple times and identified the source in different variability classes \citep{Yad16, Ban21, Maj22}. In addition, \citet{Sre20} studied the broadband spectrum of the source to estimate the mass and the spin of the system, and they concluded that GRS 1915+105 hosts a maximally rotating, comparatively higher mass BH and the X-ray binary source accreting at Super-Eddington rate. Spectro-timing properties of a new black hole transient MAXI J1348-630 have been investigated using the simultaneous/quasi-simultaneous {\it AstroSat} and {\it NICER} observations. The analysis revealed several interesting characteristics of the source, including the flip-flop behaviour, different types of QPOs and energy-dependent behaviour of variability components \citep{Jit21}. The detailed broadband spectral study using the reflection models constrained the spin and inclination of the system and the values are $> 0.97$ and $\sim 33 ^{\circ}$, respectively \citep{Mall23}. The studies mentioned above showcase the spectro-timing capabilities of {\it AstroSat} in broadband energy. 

Swift\,J1658.2$-$4242 is a low-mass black hole X$-$ray transient in
the Galactic plane and its detection was reported by
{\it Swift/BAT} on 2018 February 16 \citep{Bar18}. {\it INTEGRAL} also detected the source during the observations of the Galactic center field on 2018 February 13 \citep{GrebenevEtal2018ATel11306....1G}. The spectral hardness was typical for BHXRBs in the hard state during that period \citep{GrebenevEtal2018ATel11306....1G,GrinbergEtal2018ATel11318....1G}. The radio observations from ATCA imply that the source is a black hole X-ray binary located at a distance $>3$ kpc \citep{RussellEtal2018ATel11322....1R}. LFQPO frequencies were detected by several satellites ({\it NuSTAR, Swift, XMM-Newton}; \citet{XuEtal2018ApJ...865...18X, XuEtal2019ApJ...879...93X}, {\it AstroSat}; \citet{BeriAlta2018ATel12072....1B, JitheshEatal2019ApJ...887..101J, XiaoEtal2019JHEAp..24...30X}, {\it Insight-HXMT} and {\it NICER}; \citet{XiaoEtal2019JHEAp..24...30X}). The X-ray spectra showed high
absorption column density \citep[][and references therein]{LienEtal2018GCN.22419....1L,JitheshEatal2019ApJ...887..101J}.  Furthermore, several dips were observed in the {\it NuSTAR} light curve \citep{XuEtal2018ApJ...865...18X}, which can be a signature of high inclination ($i >65^\circ$). The joint observations from {\it NuSTAR, Swift} and {\it XMM-Newton} identified the source in the intermediate state of BHXRBs and observed a rapid flux variability in 40 s accompanied by a transient QPO with a frequency of 6--7 Hz \citep{XuEtal2019ApJ...879...93X}. The authors proposed that the accretion disc instabilities triggered at a large disk radius may be the reason for the unusual spectro-timing properties. A detailed study using the {\it Insight-HXMT}, {\it NICER} and {\it AstroSat} observations \citep{XiaoEtal2019JHEAp..24...30X} identified different timing and spectral states of the source through the hardness-intensity diagram and hardness-rms diagram. In addition, the simultaneous observations by three satellites identified the source in HIMS and detected a QPO at $\sim 1.5$ Hz frequency. The detected QPO characteristics are remarkably similar in the three energy bands considered. Extreme varieties of `flip-flop' behaviours have been observed in the source from the {\it XMM-Newton}, {\it NuSTAR}, {\it AstroSat}, {\it Swift}, {\it Insight-HXMT} and {\it INTEGRAL} observations \citep{BogensbergerEtal2020A&A...641A.101B}. During the flip-flop events, the flux was varied by $\sim 77\%$ and the flip-flop transitions occurred at random integer multiples of different fundamental periods in the two intervals of the outburst. In addition, the source exhibited changes in the power density spectrum simultaneous with the sharp changes in the flux and showed a transition between different types of QPOs (particularly type C and A) during the flip-flop. The spectral modelling of the flip-flop states revealed that the change in the inner accretion disc temperature leads to the major flux variation during the flip-flop events. 

In this work, we revisited the broadband X-ray spectral and temporal properties of Swift\,J1658.2–4242 from {\it AstroSat} observations to understand the accretion flow behaviour of the system using accretion-ejection based physical models. For this, we analyzed the simultaneous SXT and LAXPC data of Swift\,J1658.2$-$4242 and modelled the spectra with the physical model \jetcaf. In the next section, we describe the observations and the data reduction procedures used in this work. The spectral and temporal analysis results are presented in Section \S 3. In Section \S 3.4, we estimate the binary parameters from the decay of the light curve and the spectral modelling. Finally, we discuss the results and draw brief conclusions, which are presented in Section \S 4.

\section{Observation and Data Reduction}\label{obs_sec}

For this work, we used three {\it AstroSat} observations of Swift\,J1658.2–4242 and the observation details are given in Table \ref{obslog}.

\begin{table}
	\caption{Observation Log. (1) Data; (2) Observation ID; (3) date of observation; (4) modified Julian day; (5) exposure time.}

	\begin{tabular}{lcccc}
	    \hline
		Data & ObsID & Date &MJD  & Exp.\\
            &        & (in 2018)     &     & (ks)\\
	    \hline
		Data1 & T02\_004T01\_9000001910 & February 20--21 & 58170.5 &20 \\
		Data2 & T02\_011T01\_9000001940 & March 03--04 & 58180.5 &30  \\
		Data3 & T02\_020T01\_9000001990 & March 28--29 & 58205.5& 30  \\
	    \hline
			\end{tabular}
	\label{obslog}
 \end{table}
 
\subsection{Soft X-ray Telescope}
SXT \citep{Sin16, Sin17} is an imaging telescope onboard {\it AstroSat} operating in the soft X-ray energy range, 0.3--8 keV. The Level 1 Photon Counting (PC) mode data is reprocessed to obtain the cleaned Level 2 data from all orbits in the observation using the SXT pipeline software (version: AS1SXTLevel2-1.4a). The different orbit event files were merged using the SXT event merger script (based on Julia) and obtained a merged, exposure-corrected event file for the further extraction of science products. The source spectrum is extracted using the tools available in {\sc xselect}, and we used a circular region of radius 16 arcmin to extract source events from the merged event file. We generated the SXT off-axis auxiliary response file (ARF) by the {\tt sxtARFModule} tool using the on-axis ARF of version 20190608 provided by the SXT instrument team. A blank sky background spectrum provided by the SXT instrument team is used as the background spectrum in the spectral modelling. While fitting the SXT spectrum, we modified the gain of the response file using the {\sc xspec} command {\tt gain}. The tool {\tt gain} has two parameters: the slope and offset. The offset is free to vary, while the slope is fixed at 1. The spectrum in the 0.7--6 keV energy band is used for the spectral modelling.  

\subsection{Large Area X-ray Proportional Counter}
LAXPC has three units, LAXPC10, LAXPC20 and LAXPC30 operate in the 3--80 keV energy band. The three units have a total effective area of $\sim 8000~\rm cm^{2}$ and record the X-ray events with a time resolution of $10\,\rm \mu s$ \citep{Yad16, Yad16a, Ant17, AgrawalEtal2017JApA...38...30A}. We used the Event Analysis (EA) mode data of the source and reduced the Level 1 data using the LAXPC software (LaxpcSoft) to generate the Level 2 cleaned event file. Among the three units, LAXPC10 and LAXPC30 showed abnormal behaviours (gain change, low gain and gas leakage). In addition, the LAXPC 30 has no longer been operational since 2018 March 8. Thus, we used only the LAXPC20 spectrum for the spectral analysis. We modelled the LAXPC20 spectrum in the 5-25 keV energy range. We restrict our spectral analysis below 25 keV because the background dominates above 25 keV.

\section{Analysis and Results}
The spectro-temporal studies of Swift\,J1658.2-4242 have been done by several authors using different models. The authors concluded that the source is highly variable, highly absorbed with a high inclination angle \citep{XuEtal2018ApJ...865...18X}. In addition, the source showed a unique feature during the outburst (particularly in the intermediate state), namely the rapid disappearance of QPO in the high flux state (transient QPO) but in the low flux state, a significantly higher rms reported at the QPO frequency \citep{XuEtal2019ApJ...879...93X, JitheshEatal2019ApJ...887..101J, BogensbergerEtal2020A&A...641A.101B}. Though several models have been employed to understand the spectral state and QPOs separately in the past, none of the models have taken into account the accretion behavior, 
how the mass accretion rate changed, and what was the geometry of
the corona. Therefore, it is worth performing the analysis using an accretion disc-jet based model to shed further light on the aspects of the observed features. This motivates us to apply the \jetcaf\, model to this source, which is discussed in the next section.

\subsection{\jetcaf\,model}
The spectral properties of BHXRBs can be well explained using the combined disc-corona models \citep{Galeev1979ApJ...229..318G,Zdziarski2003MNRAS.342..355Z,Gilfanov2003A&A...410..217G}. Alternatively, an accretion disc based two-component advective flow \citep[TCAF;][]{ChakTitarchuk1995ApJ...455..623C} model is also available, which describes the accretion processes, the physical origin of the corona and temporal properties of the flow from the model parameters. However, this model does not take into account the signatures of jet from the system. Only recently, \jetcaf\, \citep[Jet in TCAF;][]{MondalChak2021ApJ...920...41M} model has been proposed, which includes the effect of both mass accretion and outflows in the emitted spectrum. Throughout this paper, we employed the \jetcaf\, model to fit the {\it AstroSat} spectra. 

Matter moving inwards towards the BH becomes supersonic and forms a shock in accretion \citep{Fukue1987PASJ...39..309F,Chakrabarti1989ApJ...347..365C}. Since then, many numerical simulations have been performed to verify the formation of the shock in sub-Keplerian flow \citep[][and references therein]{ChakrabartiMolteni1993ApJ...417..671C,RyuEtal1997ApJ...474..378R,GiriChak2013MNRAS.430.2836G,GarainEtal2014MNRAS.437.1329G,KimEtal2017MNRAS.472..542K}. From this shocked region to the inner sonic point is known as post-shock region. As the infalling matter gets compressed due to the geometrical effect, its optical depth and temperature also change. In this region, electrons and protons are thermally decoupled due to their different cooling timescales, which is related to their mass difference. Therefore, for this region, we solve two temperature equations, and the electron temperature is obtained as a function of optical depth \citep[cf. sec 2.4 in][]{ChakTitarchuk1995ApJ...455..623C}. The average temperature of the local plasma is estimated after deriving an accurate analytical form considering a spherical geometry of the flow. The typical value of the temperature of this region varies from a few 100 keV (in the hard spectral state) to a few keV (in the soft state). However, the temperature may become higher depending on the input parameters, which vary in multi-dimensional space. A large set of spectra were generated using a range of parameter space in \citet{Debnathetal2014} to generate a model fits file. It takes into account the maximum possible range of electron temperature and energy spectral index.

The \jetcaf\,model has three components: first, the Keplerian disc is lying at the equatorial plane which produces multicolour disc blackbody spectrum and supplies soft photons. Second, the sub-Keplerian flows above and below the Keplerian flow, which forms the dynamic corona or post-shock region. This region upscatters the soft photons from the Keplerian component as in \tcaf\, model \citep{ChakTitarchuk1995ApJ...455..623C}. As the repeated scattering of photons gains energy from the corona, the corona cools down. This gain of energy is taken care of by an enhancement factor. The third component is the mass outflow/jet implemented in \jetcaf. Considering the temperature of the post-shock region as the base, the jet temperature in the \jetcaf\, model falls down inverse to the distance away from the base of the jet. The temperature of this component varies between the other two components (the Keplerian disc and post-shock region). Therefore, soft photons from the standard disc can also get upscattered by the jet component and produce hard radiation at the shoulder of the multicolour disc blackbody peak. At the same time, the hard radiation from the hot inner region gets down scattered by the bulk motion of the jet when it passes through the jet, which is called bulk motion Comptonisation by the jet \citep{TitarchukShrader2005,MondalChak2021ApJ...920...41M}. The second component from the jet produces an excess in the spectrum above 10 keV, similar to the so-called reflection hump. The present version of the \jetcaf\, model does not consider the Iron line, which will be incorporated in the future. However, in the present analysis, the data above 6 keV is not taken into account due to calibration issues, therefore the iron line is not required.

In addition to the spectral properties, the \jetcaf\,model also explains the origin of temporal properties, e.g., the origin of QPOs, similar to the \tcaf\, model. As the matter comes closer to the BH, due to compression, it heats up, at the same time, the soft photons from the Keplerian disc get upscattered by the hot electrons, therefore cooling increases. However, both heating and cooling processes have different timescales. Once the cooling timescale matches with the heating timescale, resonance occurs and QPOs originate \citep{MolteniEtal1996ApJ...457..805M,ChakrabartiEtal2015MNRAS.452.3451C,GarainEtal2014MNRAS.437.1329G}. Thus, the model is capable of simultaneously explaining the spectral and temporal properties of BHXRBs.

The \jetcaf\, model has six parameters. These parameters are namely, (1) mass of the BH ($M_{\rm BH}$), if the mass is unknown, (2) the Keplerian or disc mass accretion rate ($\dot m_d$), (3) sub-Keplerian or halo mass accretion rate ($\dot m_h$), (4) size of the dynamic corona or the location of the shock ($X_s$) in Schwarzschild radius ($r_g=2GM_{\rm BH}/c^2$) unit, where, G and $c$ are the gravitational constant and the speed of light, respectively, (5) shock compression ratio (R=post-shock density/pre-shock density), and (6) jet collimation factor ($f_{\rm col}$=solid angle subtended by the outflow/inflow). The mass accretion rates are in the unit of Eddington rate $\dot M_{\rm Edd}$. As the matter falls in and goes out as jet/outflows, there should have some connections between accretion and ejection, which are derived in \citet{Chakrabarti1999}. It shows that the ratio of the solid angles between the outflow and inflow automatically comes to the relation, which we denoted as the jet collimation parameter ($f_{\rm col}$) in the present version of \jetcaf. However, it can be derived from the geometry of the accretion-ejection flows, which we plan to implement in the future. If the jet collimation parameter increases, the mass outflow rate, optical depth and the number of scattering increase. This makes the spectrum harder. Apart from the above parameters, the model requires a scale factor, the  normalization (norm), to match the theoretical spectra to the observed one. Therefore, it should not vary drastically from observation to observation for a single source in a single outburst. Several works \citep[][and others]{MollaEtal2016MNRAS.460.3163M, MollaEtal2017ApJ...834...88M} have discussed this issue and estimated the mass of the central BH using the constant normalization method. In this model, the norm, which for simplicity could be written as $R_z^2/(4 \pi D^2)\sin{i}$, where, `$R_z$' is the effective height of the Keplerian component in km at the pre-shock region, `D' is source distance (in 10 kpc unit) and `$i$' is disc inclination angle with the line of sight is also a variable. However, since the value of `$R_z$' is unknown, we leave it as a free parameter to be determined from the best-fitting value. This is very similar to the estimation of Rin, the inner edge of the disk, obtained from the normalization of the {\diskbb} model fit. 

The \jetcaf\,model is implemented as a local additive model\footnote{https://heasarc.gsfc.nasa.gov/xanadu/xspec/manual/node100.html} in {\sc xspec} \citep[version 12.12.0;][]{Arnaud1996ASPC..101...17A} for the spectral fitting. Along with the \jetcaf\,model, we have used \tbabs\, \citep{Wilms2000ApJ...542..914W}, partial covering absorption (\pcfabs), \textcolor{black}{and partial covering absorption by partially ionized plasma \citep[\zxipcf;][]{ReevesEtal2008MNRAS.385L.108R}}. Using these model components, we fitted the broadband {\it AstroSat} spectra in the 0.7--25 keV energy band.

\begin{figure*}
\begin{center}
\hspace{-2.95cm}
\includegraphics[height=5truecm,angle=0]{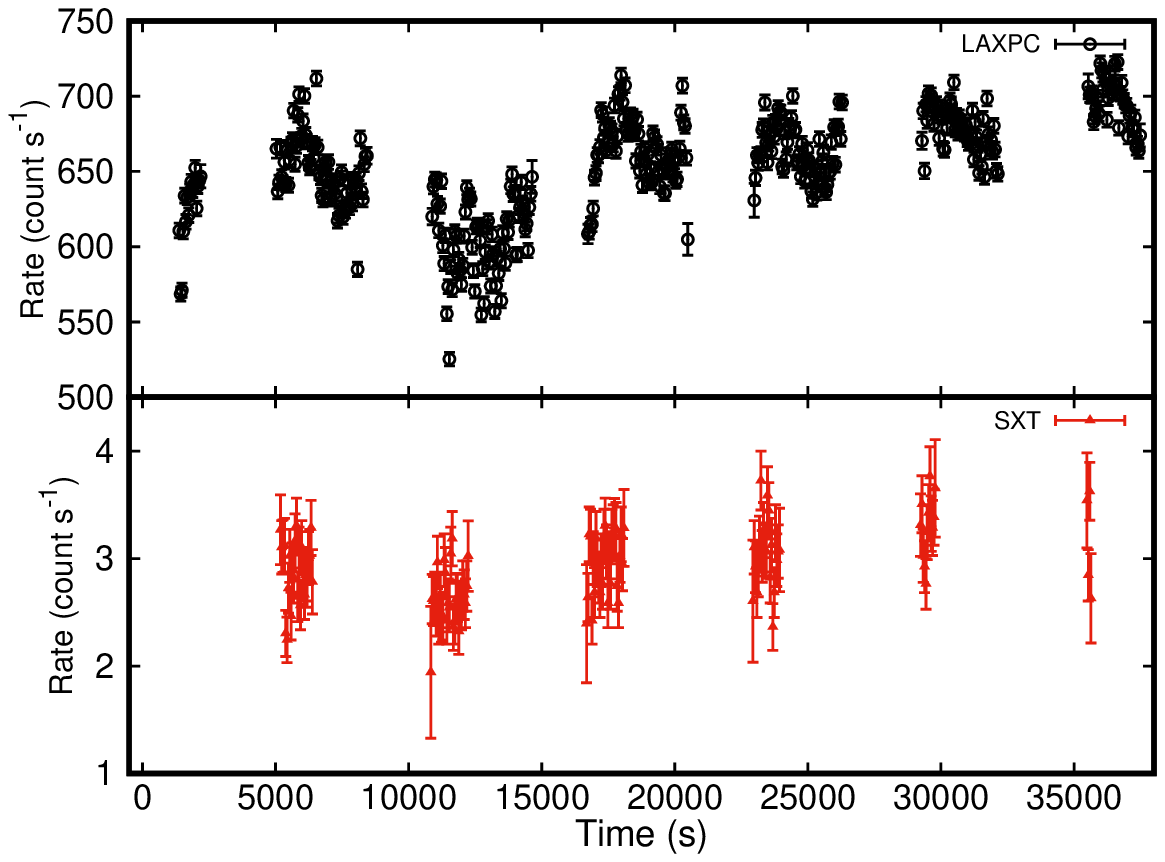}
\hspace{-0.6cm}
\includegraphics[height=5truecm,angle=0]{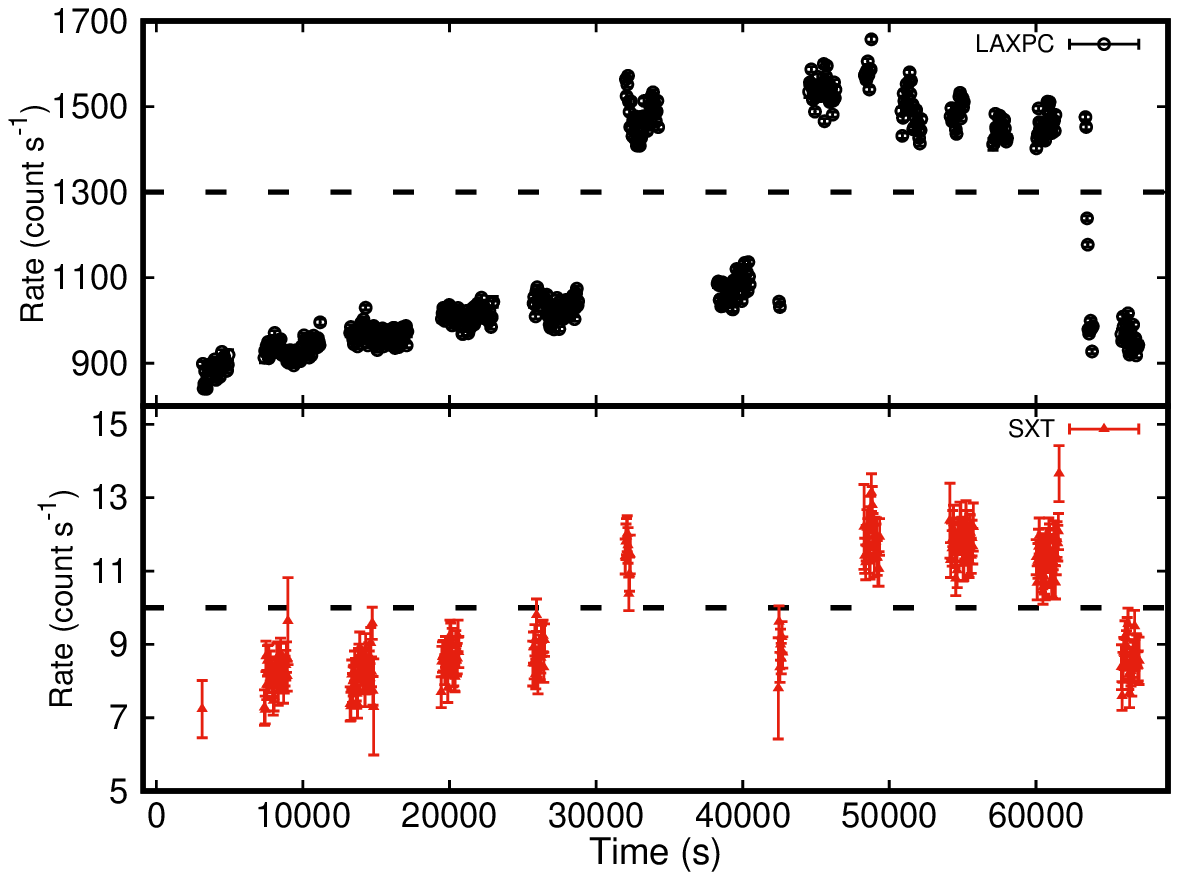}
\hspace{-0.6cm}
\includegraphics[height=5truecm,angle=0]{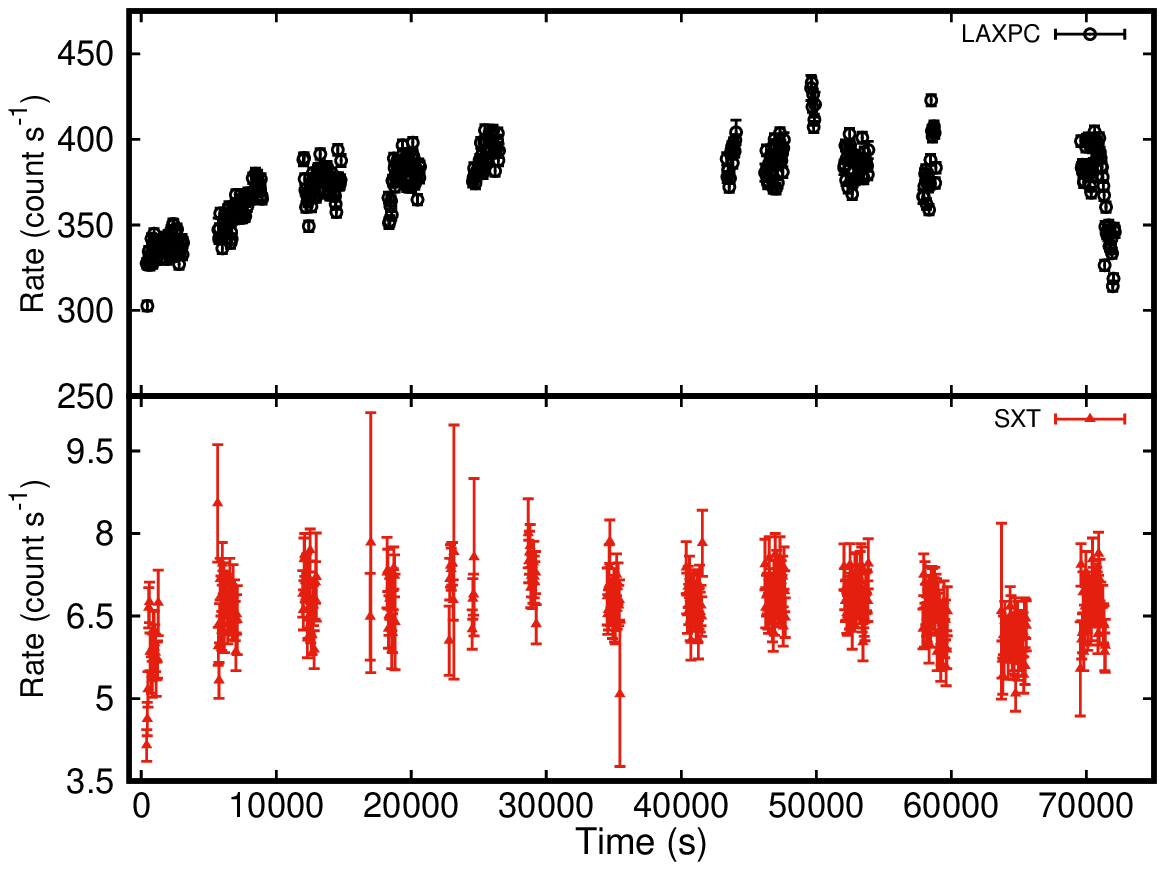}
\hspace{-3.0cm}
\caption{ {\it AstroSat} LAXPC and SXT light curves of Swift\,J1658.2–4242 from Data1 (left), Data2 (middle) and Data3 (right). The top and bottom panels in each plot represent the LAXPC (3--50 keV) and SXT (0.3--8 keV) light curves, respectively. The time bin size for LAXPC and SXT light curves is 50\,s. We used LAXPC10 and LAXPC20 for the light curve extraction in Data1 and Data2, while for Data3, LAXPC20 alone was used. The black dashed horizontal lines in Data2 represent the demarcation of the two flux levels.}
\label{laxpc_sxt_lc}
\end{center}
\end{figure*}

\subsection{Spectral fitting}
As mentioned in \S \ref{obs_sec}, we used three {\it AstroSat} observations of Swift\,J1658.2–4242 for this study and the LAXPC and SXT light curves are given in Figure \ref{laxpc_sxt_lc}. In all observations, the source intensity varies in both SXT and LAXPC. Particularly in Data2, the source exhibits a significant flux variability, where the source intensity gradually increases in the initial period of the observation, then it changes abruptly to a high intensity level. After spending a few kilo seconds in the high intensity level, the source comes back to low intensity state and goes into the high intensity quickly once again. In the last part of the observation, the source makes a transition from the high intensity to the lower intensity level (see the middle panel of Figure \ref{laxpc_sxt_lc}). This variability behaviour is similar to flip-flop behaviour observed in BHXRBs \citep{Miy91, Tak97, Sri12, SritamEtal2016ApJ...823...67S}. In addition, the source exhibited different variability features in the {\it AstroSat} LAXPC data. In particular, a low-frequency QPO at $\sim 1.56$\,Hz drifted to $\sim 1.74$\,Hz during the course of the first observation (Data1), while a transient QPO associated with the flip-flop behaviour has been observed in the Data2. In Data3, a QPO detected at $\sim 3.98$ Hz \citep[see][]{JitheshEatal2019ApJ...887..101J}. The main motivation of \citet{JitheshEatal2019ApJ...887..101J} was to understand the energy-dependent variability properties and the spectral state. The single-zone stochastic propagation model successfully explained the energy-dependent variability properties of the source and shed light on the mechanism responsible for the variability. In the broadband spectral analysis, they used the phenomenological model to understand the spectral state mainly. This motivates us to do detailed spectral modelling with a complex physical model like \jetcaf\, to understand the observed properties of the source.

\begin{figure}
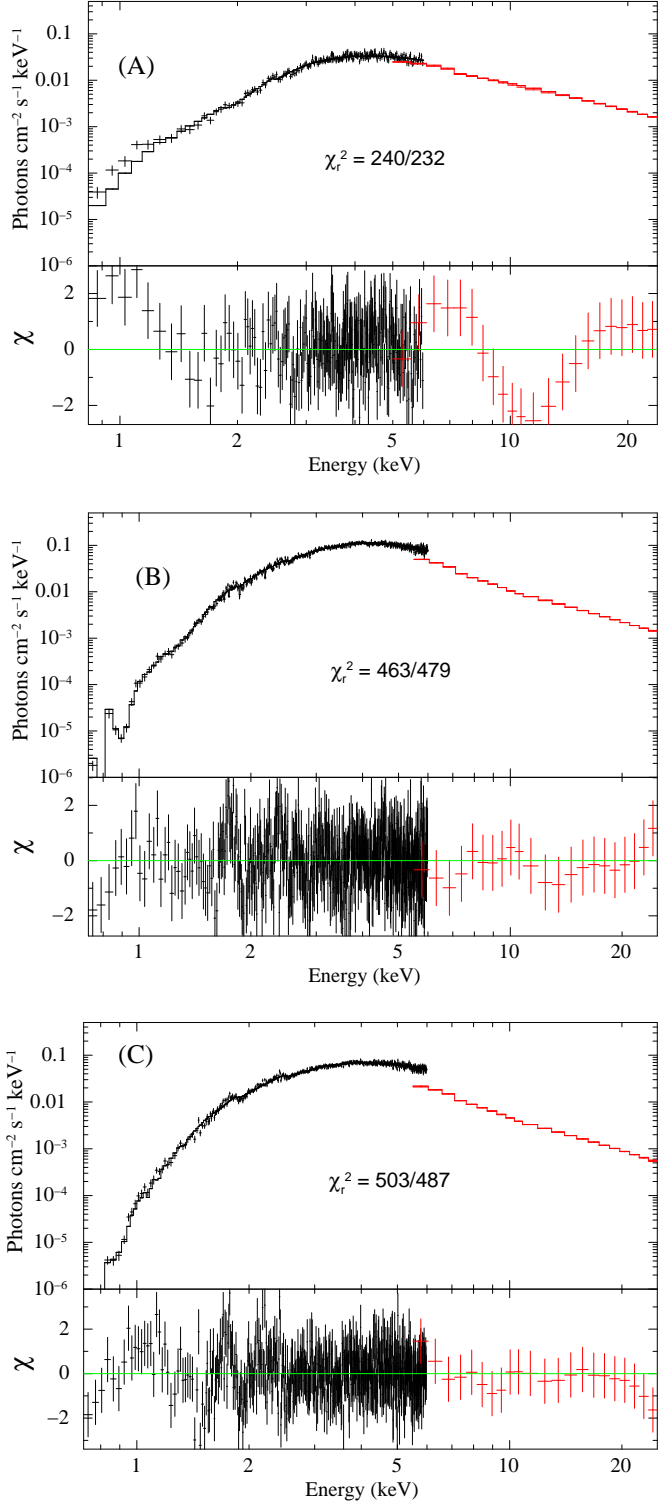

\centering{ 
\includegraphics[height=9truecm,angle=270]{Figures/O1-r2-tb-pcf-zxi-jet-fcol.eps}
\includegraphics[height=9truecm,angle=270]{Figures/O2-r2-tb-pcf-zxi-jet-fcol.eps} %\\
\includegraphics[height=9truecm,angle=270]{Figures/O3-r2-tb-pcf-zxi-jet-fcol.eps}}
\caption {The SXT (black points) and LAXPC (red points) unfolded spectra from Data1 (top panel), Data2 (middle panel), Data3 (bottom panel) are fitted with \textcolor{black}{the model M2}.}
\label{fig:SpecJetcaf}
\end{figure}

\textcolor{black}{We fitted the broadband spectra of these {\it AstroSat} observations using the \jetcaf\,model along with different absorption models available in the literature. Our fitting uses different model combinations to estimate the physical accretion parameters and the significance of absorption, as the source is highly absorbed. 
First, we fit the data using model \tbabs*\pcfabs*\jetcaf\,(M1), where the Galactic absorption column density ($N_{\rm HG}$) was fixed to its known value of $1.83\times 10^{22}$ cm$^{-2}$ \citep{KalberlaEtal2005A&A...440..775K}. This gives a poor fit with reduced $\chi^2$ ($\chi_r^2$) $\gtrsim$2. Therefore, for this model fit, we could not estimate errors for model parameters. The parameter values are given in the top panel of \autoref{table:jetcafResults}. In the next step of the analysis, we let the $N_{\rm HG}$ parameter vary, and obtained an improved fit ($\chi_r^2 \sim 0.9-1.1$ for higher $N_{\rm HG}$ values. However, the Galactic absorption neither varies significantly from observation to observation in the same outburst nor can be much higher than (by a factor of few) its known line of sight value between the source and the observer. Nonetheless, this model fit gives an indication that some intrinsic absorption is there, which needs to be taken care of by using some physically motivated absorption models. A similar conclusion was drawn by \citet{XuEtal2019ApJ...879...93X}, however, by considering a single component absorption model with $N_{\rm H}$ value an order of magnitude higher than known $N_{\rm HG}$ \citep{KalberlaEtal2005A&A...440..775K}.} 

\textcolor{black}{Moreover, the \jetcaf\, model parameters show that the disc mass accretion rate ($\dot m_{\rm d}$) of the source is close to the Eddington accretion rate and the mass outflow parameter ($f_{\rm col}$) was also moderately high. For such high accretion state source, it might be possible that soft photons from the disc may get Comptonised by some warm and optically thick corona which can be originated from the disc wind or outflows \citep[][and references therein]{DoneEtal2018MNRAS.473..838D}. Therefore, to consider such effects, we added a partially ionised plasma (pip) model (\zxipcf) to M1 and the model \tbabs*\pcfabs*\zxipcf*\jetcaf\, is designated as M2. During fitting with M2, we kept N$_{\rm HG}$ to the known Galactic value. The best fit is achieved using model M2 with $\chi_r^2$ values of $\sim 1.0$ for all data sets. These results and fit statistics are given in the second panel of \autoref{table:jetcafResults}. It can be seen that the column density of \pcfabs\,model (N$_{\rm Hpcf}$) varies between $8.5-12.7\times10^{22}$ cm$^{-2}$ and the partial covering factor (C$_f^{pcf}$) was fluctuating between 0.58 and 0.95. The column density of \zxipcf\,model (N$_{\rm Hpip}$) varies noticeably between $3.9-8.3\times10^{22}$ cm$^{-2}$ with partial covering factor (C$_f^{pip}$) $>0.9$ and pegged to its upper limit. The ionisation parameter ($\xi$) is varied between 2.5-25 erg cm s$^{-1}$ and is relatively low. The above model parameters and their covering coefficients imply that the source is highly absorbed, and significant intrinsic absorption is there, which can be due to wind/outflow from the highly accreting disc. The model M2 is fitted to the data and is shown in \autoref{fig:SpecJetcaf}. Panels A, B, and C in \autoref{fig:SpecJetcaf} denote the Data1, 2, and 3, respectively. The \jetcaf\,model continuum and different absorption models are adequate to fit the present data sets.}

\textcolor{black}{During the fitting with different model combinations, the mass of the black hole (M$_{\rm BH}$) in \jetcaf\, model is used as a free parameter, since we do not know its value beforehand. As expected, the mass should not change from epoch to epoch, but rather remain constant. This is consistent with the yielded average mass value from X-ray spectral fitting using M2, which is $\sim 14$ M$_\odot$ (see \autoref{table:jetcafResults}). The disc mass accretion rate ($\dot m_d$) in \jetcaf\, model (in M2) varies between 0.9 to 1.04 $\dot M_{\rm Edd}$, whereas the halo accretion rate ($\dot m_h$) is relatively low and varies in the range of 0.15--0.24 $\dot M_{\rm Edd}$.  
We have noticed a significant change in the size of dynamic corona ($X_{\rm s}$) from $\sim 34~r_g$ to $\sim 65~r_g$  during the outburst period, and anti-correlates with $\dot m_d$. The shock compression ratio (R) variation was significant, from 5.6 to 6.6 in Data1, 2, and 3. The jet collimation factor ($f_{\rm col}$) was low ($\sim 0.01$) during the first observation and increased to $\sim 0.4$ in the second and third observations. This is also reflected in both the column densities, that have increased in Data2 and 3 compared to Data1. This implies that jet/mass outflow from the inner region of the disc was present, changed the shape of the continuum. We note that the \jetcaf\,model {\it norm} varies between 8--11 for all observations. For Data1, the {\it norm} is 3. As we have only three observations, we can not apply the constant normalization technique as used in \citet{MollaEtal2016MNRAS.460.3163M,MollaEtal2017ApJ...834...88M}. However, naturally, a consistent BH mass is getting from the best fit. The best-fit spectral parameters are provided in \autoref{table:jetcafResults}. All errors are estimated in the 90\% confidence level using {\it err}\footnote{https://heasarc.gsfc.nasa.gov/xanadu/xspec/manual/node79.html  https://heasarc.gsfc.nasa.gov/xanadu/xspec/manual/node28.html} command. Interestingly, in both model combinations, \jetcaf\, model parameters show a similar trend of parameter variations and the parameter values are quite similar. This concludes that the \jetcaf\, model parameter estimation is robust. We have computed the best-fit \jetcaf\,model spectra without absorption using the source code for the best-fitted model M2 parameters, which is shown in \autoref{fig:SpecNH0}.} 

\begin{figure}
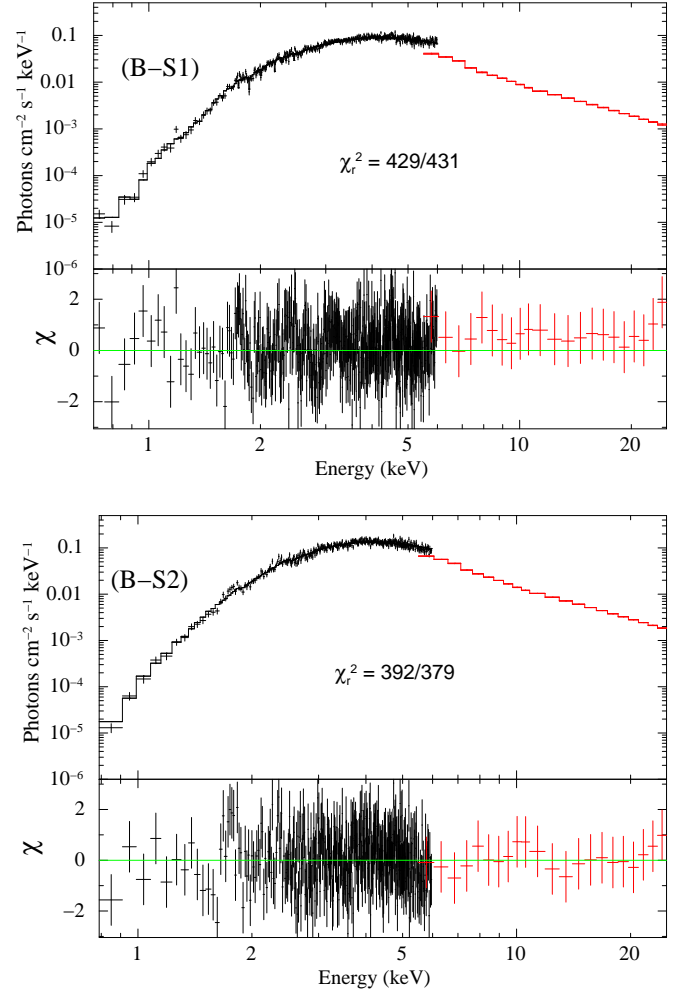

\centering{ 
\includegraphics[height=9truecm,angle=270]{Figures/O2-s1-r2-tb-pcf-zxi-jet-fcol.eps}
\includegraphics[height=9truecm,angle=270]{Figures/O2-s2-r2-tb-pcf-zxi-jet-fcol.eps}}
\caption {The flux-resolved spectral analysis of Data2 with \textcolor{black}{the model M2}. The top and bottom panels represent the spectral fit for the low and high flux levels, respectively.} 
\label{fig:Seg2Spec}
\end{figure}

%========================================================
\textcolor{black}{The source exhibited flip-flop behaviour in Data2. To understand this variability behaviour, we further extracted the flux-resolved spectra from Data2 and performed the spectral modelling using all three models discussed above. During this fitting, we fixed the mass of the black hole to the values obtained from the Data2 and freed the other parameters. The best-fit spectral parameters from the flux-resolved spectral analysis using model M2 are given in \autoref{table:jetcafResults} (as Seg 1 and 2). The two segments are also fitted well with model M2, with $\chi_r^2 \sim$ 1. The spectral fits are shown in \autoref{fig:Seg2Spec}. In this analysis, both mass accretion rates ($\dot m_{\rm d}$ and $\dot m_{\rm h}$) increase as the flux increases from low to high flux states, while the size of the corona and shock compression ratio decrease during the transition. In addition, we notice that $\dot m_d$ has increased on the same day, which has changed the size of the dynamic corona from $\sim 40~r_g$ to $\sim 37~r_g$. The $f_{\rm col}$ parameter is more or less the same, while the R parameter increases its value from 5.7 to 6.1.  The {\it norm} of \jetcaf\,model is $\sim 10$ for both segments. The N$_{\rm Hpcf}$ value has increased from 6 to 8 $\times 10^{22}$ cm$^{-2}$ and the N$_{\rm Hpip}$ has decreased from 14 to 8 $\times 10^{22}$ cm$^{-2}$ along with a decrease in $\xi$ from 63 to 13 erg cm s$^{-1}$. In the next section, we estimate the QPO frequency using the best-fit shock parameters of the \jetcaf\,model in M2.}

%=======================

\begin{table*}
%\hspace{-12.0cm}
%\scriptsize
%\small
\centering
\caption{The broadband spectral parameters of Swift\,J1658.2$-$4242 when fitted with models M1, and M2. \textcolor{black}{Here, N$_{\rm Hpcf}$, C$_f^{\rm pcf}$, N$_{\rm Hpip}$, and C$_{\rm f}^{\rm pip}$ are hydrogen column densities and dimensionless covering fractions for \pcfabs\, and \zxipcf\, models, respectively and $\xi$ is the ionization parameter for the \zxipcf\,model. The C$_{\rm f}^{\rm pip}$ parameter obtained from the fit is high $> 0.9$ and pegged at the upper limit for all observations. The $\dot m_d$, $\dot m_h$, $X_s$, $R$, and $f_{\rm col}$ are the disc and halo mass accretion rates, location of the shock or size of the corona, shock compression ratio, and jet collimation factor respectively.} $^f$ denotes the values are frozen. The Galactic absorption (N$_{HG}$) value is \textcolor{black}{frozen to $1.83\times 10^{22}$ cm$^{-2}$ for all observations. For poor fits with $\chi_r^2 \gtrsim 2$, errors of the model M1 parameters are not estimated.}} \label{table:jetcafResults}
\resizebox{\textwidth}{!}{\begin{tabular}{ccccccccccccccccc}
\hline
Model&Data&Seg&$N_{\rm Hpcf}$&C$_f^{\rm pcf}$&$N_{\rm Hpip}$&$\log(\xi_{\rm pip})$&M$_{\rm BH}$&$\dot m_{\rm d}$  & $\dot m_{\rm h}$ & $X_{\rm s}$ & R &f$_{\rm col}$&$\chi^2_{r}/dof$ \\
& & &$(10^{22}$ cm$^{-2})$& &$(10^{22}$ cm$^{-2})$&(erg cm s$^{-1}$) &(M$_\odot$)&$(\dot M_{\rm Edd})$&$(\dot M_{\rm Edd})$&$(r_{\rm g})$& & & & &  \\
\hline
%&&&&&&& M1&&&&&&\\
%\hline
&1 & - &$8.1^{}_{}$&$0.95^{}_{}$&$-$&$-$&$13.0^{}_{}$&$0.91^{}_{}$&$0.14^{}_{}$&$65.0^{}_{}$&$5.1^{}_{}$& $<0.01$&2.4/235\\
&2 & - &$7.3^{}_{}$&$0.95^{}_{}$&$-$&$-$&$13.8^{}_{}$&$0.92^{}_{}$&$0.20^{}_{}$&$34.2^{}_{}$&$5.6^{}_{}$&$0.40^{}_{}$&3.4/482\\
M1 &  & 1 &$7.6^{}_{}$&$0.95^{}_{}$&$-$&$-$&$^f13.8$&$1.05^{}_{}$&$0.24^{}_{}$&$38.4^{}_{}$&$6.0^{}_{}$&$0.39^{}_{}$ &1.8/434\\
   &  & 2 &$10.1^{}_{}$&$1.0^{}_{}$&$-$&$-$&$^f13.8$&$1.16^{}_{}$&$0.21^{}_{}$&$41.2^{}_{}$&$5.3^{}_{}$&$0.32^{}_{}$&3.3/382\\
&3 & - &$7.9^{}_{}$&$0.95^{}_{}$&$-$&$-$&$14.0^{}_{}$&$1.05^{}_{}$&$0.27^{}_{}$&$32.1^{}_{}$&$6.2^{}_{}$&$0.33^{}_{}$&1.9/490\\
%\hline
%&&&&&&& M2&&&&&&\\
\hline
&1 & - &$8.5^{+0.9}_{-0.7}$&$0.95^{+0.07}_{-0.04}$&$3.9^{+0.2}_{-0.1}$&$0.4^{+0.1}_{-0.1}$&$13.5^{+0.5}_{-0.4}$&$0.90^{+0.02}_{-0.01}$&$0.15^{+0.01}_{-0.01}$&$64.9^{+3.9}_{-3.1}$&$5.6^{+0.2}_{-0.1}$& $<0.01$ &1.0/232\\
&2 & - &$9.2^{+1.0}_{-0.9}$&$0.79^{+0.05}_{-0.04}$&$8.3^{+0.5}_{-0.4}$&$1.4^{+0.2}_{-0.1}$&$13.6^{+0.6}_{-0.5}$&$0.93^{+0.04}_{-0.04}$&$0.20^{+0.01}_{-0.01}$&$34.5^{+2.0}_{-1.5}$&$5.7^{+0.3}_{-0.2}$&$0.39^{+0.05}_{-0.03}$ &1.0/479\\
 M2&  & 1 &$6.0^{+0.4}_{-0.3}$&$0.72^{+0.06}_{-0.06}$&$14.4^{+1.1}_{-0.9}$&$1.8^{+0.3}_{-0.1}$&$^f13.6$&$1.07^{+0.03}_{-0.02}$&$0.22^{+0.02}_{-0.01}$&$40.1^{+2.2}_{-1.9}$&$5.7^{+0.4}_{-0.4}$&$0.33^{+0.04}_{-0.03}$ &1.0/431\\
 & & 2 &$7.8^{+0.9}_{-0.9}$&$0.79^{+0.06}_{-0.04}$&$7.8^{+1.0}_{-0.7}$&$1.1^{+0.2}_{-0.2}$&$^f13.6$&$1.09^{+0.04}_{-0.03}$&$0.25^{+0.02}_{-0.01}$&$37.4^{+2.1}_{-2.1}$&$6.1^{+0.5}_{-0.4}$&$0.32^{+0.02}_{-0.01}$ &1.0/379\\
&3 & - &$12.7^{+1.0}_{-0.8}$&$0.58^{+0.10}_{-0.07}$&$7.3^{+0.9}_{-0.7}$&$0.7^{+0.1}_{-0.1}$&$14.2^{+0.7}_{-0.5}$&$1.04^{+0.03}_{-0.03}$&$0.24^{+0.01}_{-0.01}$&$41.1^{+2.8}_{-2.4}$&$6.6^{+0.4}_{-0.2}$& $0.35^{+0.03}_{-0.03}$ &1.0/487\\
\hline
\end{tabular}\label{tab:TCAFResults}  }
\end{table*}

\subsection{QPO frequency from model parameters} \label{sec:qpoEstimate}
The QPO frequencies can be estimated using the shock oscillation model \citep{MolteniEtal1996ApJ...457..805M,chakrabartiEtal2008A&A...489L..41C} as below, 
\begin{equation}
   \nu_{\rm QPO}= \frac{c}{r_{\rm g}\; M_{\rm BH}\; R X_{\rm s}^{1.5}}.
\end{equation}

For the three observations, the computed QPO frequencies are 2.53, 6.37, and 4.04 Hz, respectively. The estimated QPO frequencies, the observed QPO frequencies and the Q-factor of the QPOs are given in \autoref{tab:QPOmass}. The estimated QPO values are consistent with the observed QPOs. The $\nu_{\rm QPO}$ values correlate with the $\dot m_d$, where the highest and lowest QPOs are observed during the highest and lowest accretion rate periods. Apart from the QPOs, the model can explain the broadband noise variability. In the case of QPOs, we considered that the shock is axisymmetric. However, it is not necessarily always, in a complicated disc, shock can be non-axisymmetric due to some perturbations \citep[see][for recent simulation]{ChakrabartiWiita1993A&A...271..216C,GarainKim2022arXiv221208310G}. Therefore, the spiral shock arms may deform and reform again, which can produce broad and multi-peaked power density spectra. Thus, it is also possible to explain broadband noise in the light of \jetcaf\, model after implementing non-axisymmetric shock.

\begin{table}
\small
    \centering
    \caption{Estimated QPO frequencies \textcolor{black}{using M2 model parameters} against those reported by \citet{JitheshEatal2019ApJ...887..101J} }
    \begin{tabular}{c c c c}
    \hline
    Data &Estimated $\nu_{\rm QPO}$ &Observed $\nu_{\rm QPO}$ & Q \\
%    &  &   \\  
    &(Hz) &(Hz) & \\
    \hline
    1 &$2.53^{+0.25}_{-0.19}$ &$1.57^{+0.01}_{-0.01}$& 8.60\\
       &$--$ &$1.71^{+0.04}_{-0.04}$& 4.39\\
    2 &$6.37^{+0.62}_{-0.46}$ &$6.57^{+0.05}_{-0.05}$& 3.23\\
    3 &$4.04^{+0.43}_{-0.34}$ &$3.97^{+0.04}_{-0.02}$& 8.30\\
   \hline
    \end{tabular}
    \label{tab:QPOmass}
\end{table}

\subsection{Estimating Binary Parameters} \label{sec:binary}
\begin{figure}
\centering{ 
\includegraphics[height=8truecm,angle=0]{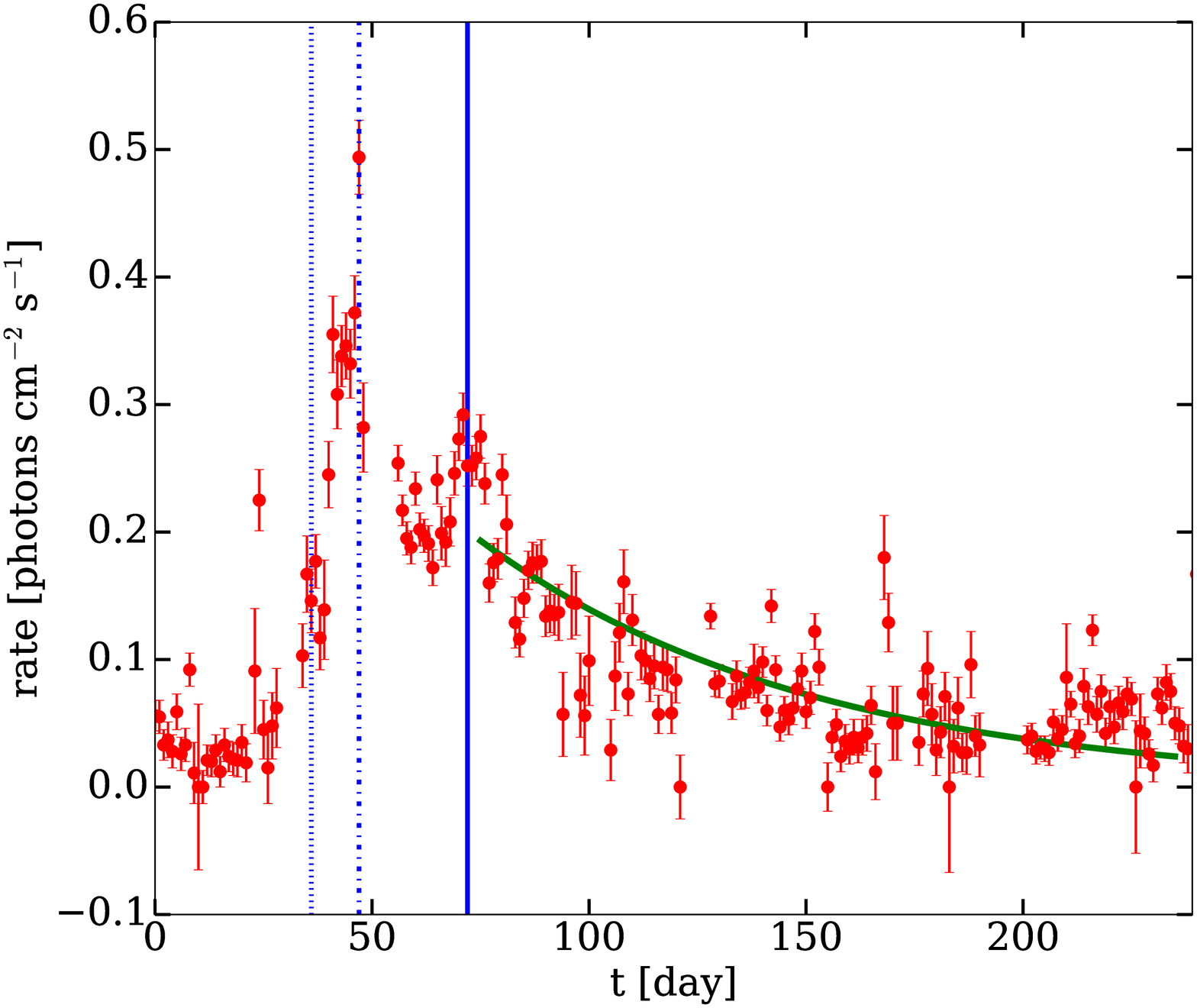}}
\caption {The 1-day binned {\it MAXI} light curve of Swift J1658.2–4242 in the 2–20 keV energy band over the
period 2018 January 14 to October 9. The day ``0" represents MJD 58132.5. The vertical dashed, dot-dashed, and solid blue lines correspond to three {\it AstroSat} observations considered in this work. The solid green line represents the exponential decay function fitted to the light curve.}
\label{fig:lc_fit}
\end{figure}

Decay of X-ray light curve is common in BHXRBs with irradiated or unirradiated discs and they generally show three decay phases in the long-term light curve during the outburst evolution \citep[][and references therein]{KingRitter1998MNRAS.293L..42K,King1998MNRAS.296L..45K,DubusEtal1999MNRAS.303..139D,FrankBook2002}. The first two decay profiles follow the exponential, while the third one has a linear profile \citep{KingRitter1998MNRAS.293L..42K}. It should be noted that the third phase of decay might not always be observed as it appears much later in the outburst. Later, \citet{Mondal2020AdSpR..65..693M} revisited these decay profiles by parameterizing $\alpha$ and combined the mass and orbital period of the companion star to understand the effect of the binary parameters. It is also noted that the binary parameters can be estimated from the same solution. \autoref{fig:lc_fit} shows the 1-day binned {\it MAXI} light curve of Swift\,J1658.2–4242 in the 2–20 keV energy band over the period 2018 January 14 to October 9. The light curve clearly shows two decays, the first one starts between Observation 2 and 3, and the second decay starts later after Observation 3. Both decay profiles more or less follow the exponential function as explained in \citet{KingRitter1998MNRAS.293L..42K} for a given viscosity parameter ($\alpha$) using disc irradiation. The first decay of the source, though it shows a sharp change, does not have sufficient data points at the slope of the decay. Hence, we did not fit this phase. As expected, this phase can provide a different decay (viscosity) timescale as the two decay phases have different physical origins. The third phase of decay was not observed in the light curve. Thus, we considered the second decay phase for the fitting, which is much clearer and lasted for a longer time. The solid green line in \autoref{fig:lc_fit} represents the best-fit exponential decay function, $f(t)=A exp(-t/t_v$) fitted to the light curve. Here, $t_v$ is the light curve decay timescale or the viscous timescale and A is the normalization. The fitted function gives A = $0.51\pm0.05$ photons cm$^{-2}$ s$^{-1}$ and $\tau=76.7\pm4.9$ day with reduced $\chi^2=4.96$ for 131 degrees of freedom. After getting the $t_v$, we estimate the binary parameters as discussed below.

The amount of matter accreting from the companion star falling into a black hole may depend on the companion properties and binary separation or orbital period of the companion star. Therefore, by fitting the spectrum and the decay profile of the light curve, it might be possible to put some light on the binary parameters. To do that, we considered the unirradiated disc mass accretion rate and the decay profile of the light curve. The effective disc temperature can be written as \citep{ShakuraSunyaev1973A&A....24..337S},
\begin{equation}
    T=\left(\frac{3 G M_{\rm BH} \dot M_d}{8\pi \sigma r_{\rm out}^3}\right)^{1/4},
\end{equation}

% \begin{equation}
%     T^4=\left(\frac{\eta \dot M_d c^2 (1-\beta)}{4\pi \sigma r_{\rm out}^2}\right) \left(\frac{H}{r_{\rm out}}\right)^2 \left[\frac{\text{d}\ln H}{\text{d}\ln r_{\rm out}}-1\right],
% \end{equation}
where G, $\dot M_d$, $\sigma$, and $r_{\rm out}$ are the gravitational constant, disc mass accretion rate in g/s unit, Stefan-Boltzman constant, and outer radius of the disc, respectively. This gives the sound speed at the radius for a given temperature (T) as
\begin{equation}
    c_s=\left(\frac{\gamma k T}{\mu m_p}\right)^{1/2}.
\end{equation}
Here, $\gamma$ (= 5/3), $k$, $\mu$ (= 0.6), and $m_p$ are the adiabatic index of the flow, Boltzman constant, mean molecular weight, and the mass of the proton, respectively. We can write the height of the disc as,

\begin{equation}
    H=c_s\left(\frac{r_{\rm out}^3}{G M_{\rm BH}}\right)^{1/2}.
\end{equation}

Once we have all these quantities, the viscous timescale can be written from \citet{ShakuraSunyaev1973A&A....24..337S},

\begin{equation}
    t_v=\frac{r_{\rm out}^2}{\alpha c_s H},
\end{equation}
here, $\alpha$ is the viscosity parameter, for which we have chosen a typical value $\sim 0.1$ \citep{KingEtal2007MNRAS.376.1740K} for the rest of the analysis.

After doing some simple algebra, $r_{\rm out}$ can be written as,
\begin{equation}
r_{\rm out}=6.15\times 10^{8}~\text{cm}~~\dot M_{d}^{1/5} M_{\rm BH}^{-1/5} t_v^{4/5} \alpha^{4/5}.
\end{equation}

Considering $M_{\rm BH}$, $\dot M_d$ (in \autoref{tab:TCAFResults}), and $t_v$ from spectral and light curve fitting, the $r_{\rm out}$ value comes out to be $2.8\times 10^{10}$ cm. The standard assumption regarding the outer accretion disc radius is that the disc is tidally truncated at about 80\% \citep[see][]{Cannizzo1994ApJ...435..389C} the Roche Lobe radius ($R_L$) of the primary. This gives $R_L \sim 3.5\times 10^{10}$ cm. Now, the binary separation ($a$) can be estimated using the relation \citep{Eggleton1983ApJ...268..368E},

\begin{equation}
R_L=\frac{0.49 a q^{2/3}}{0.6 q^{2/3}+\ln[1+q^{1/3}]},
\end{equation}

here, $q$ is the binary mass ratio (donor mass/accretor mass; $M_d/M_{\rm BH}$). The optical observation of this binary system has inferred that the companion might be a mid or late K-type star \citep{BahramianEtal2018ATel11342....1B} or an evolved star. Therefore, the mass of the donor can lie between 0.5 and 0.8 $M_\odot$. Considering the lower and upper $M_d$ values, the upper and lower limit of $a$ come out to be $a_u=2.3\times 10^{11}$ cm and $a_l=2.0\times 10^{11}$ cm.  

The orbital period ($P_{\rm orb}$) of the donor star can be estimated using Kepler's third law,
\begin{equation}
P_{\rm orb}(hr)=6.83 a^{3/2} \left(1+q\right)^{-1/2} \left(\frac{M_{\rm BH}}{M_\odot}\right)^{-1/2}.
\end{equation}
The $a_u$ and $a_l$ values give the upper and lower limits of $P_{\rm orb}$ come out as 4.4 hr and 3.6 hr. Therefore, our estimated possible $P_{\rm orb}$ value is $4.0\pm0.4$ for a reasonable $\alpha$-parameter value of 0.1. 

\section{Discussion and Conclusion}

 In this paper, we have presented the spectral and temporal analysis of the black hole transient Swift\,J1658.2--4242 using the SXT and LAXPC onboard {\it AstroSat}. \textcolor{black}{The data is fitted using the \jetcaf\, model along with different absorption model components to understand the spectral and temporal properties of the source. The estimated absorption column densities for different absorption models are high and in the range of $\sim 6-13 \times 10^{22}$ cm$^{-2}$ for \pcfabs\, and $\sim 4-14 \times 10^{22}$ cm$^{-2}$ for \zxipcf\,models, while the N$_H$ for \tbabs\,model was fixed to the Galactic value. 
 The higher value of $N_H$ suggests that the neutral absorption is probably intrinsic to the source, which might be originated from the significant wind or outflow from the disc. It is believed that mass accretion close to or higher than Eddington rate may launch winds \citep[][and references therein]{Lipunova1999AstL...25..508L,DoneEtal2018MNRAS.473..838D}, which might be the case for this source. It is further supported by the estimated disc mass accretion rate.} 

The \jetcaf\, model provides the mass accretion rate and the accretion geometry of the system directly. \textcolor{black}{The best-fit disc mass accretion rate is high and close to the Eddington rate in these observations. The jet/outflow collimation factor for Data 2 and 3 is also moderately high $\sim 0.4$, which has modified the shape of the continuum spectra. However, the low or negligibly small value in Data1 says that outflow was mild or absent during this epoch. The shock location or corona size reduced from 65 to 35 $r_g$ during the observation period due to the change in mass accretion rates. The smaller corona size for a higher accretion rate and vice-versa are consistent with the cooling rate due to the scattering of soft photons by hot electrons in the corona. Moreover, the disc mass accretion rate is always higher than the hot flow or halo mass accretion rate along with a bigger size of the corona implies that the source belongs to the intermediate to soft spectral state of BHXRBs.} 

Several works in the literature, which studied months or weeks of continuous observations of several black hole candidates in the framework of \tcaf\, model \citep[][and references therein]{MondalEtal2014ApJ...786....4M, DebnathElal2015MNRAS.447.1984D, MollaEtal2016MNRAS.460.3163M, MollaEtal2017ApJ...834...88M}, concluded that, during the rising phase of the outburst, the halo rate is high and slowly goes down, and the disc rate goes up. At the same time, shock moves inward. However, in the declining phase, the scenario is the opposite. As we do not have continuous monitoring observations for weeks or months, we can not verify those profiles. However, the basic relations between parameters in our study agree with previous studies. \textcolor{black}{In addition, different model combinations fit to the data show that \jetcaf\,model parameters do not vary from one to the other. This suggests that the accretion flow parameters estimated using \jetcaf\,model are robust.}

The QPO frequencies from all three epochs are estimated by considering the shock oscillation model and using the model fitted shock location and compression ratio parameters. The computed QPO frequencies are consistent with the observed frequencies (see \autoref{tab:QPOmass}). The $\nu_{\rm QPO}$ values are also correlated with the $X_s$, where the highest frequency QPO ($\sim 6.6$ Hz) observed at the lowest shock location values and vice-versa, indicating that the cooling may take a role in this change. The increase in $\dot m_d$ generates more soft photons in the disc, which are upscattered by the hot corona and gain energy, therefore cooling rate increases \citep[][and references therein]{MondalEtal2017ApJ...850...47M}. Thus, the corona size ($X_s$) decreases and $\nu_{QPO}$ increases. The Q-factor of the QPO correlates with the geometry of the corona, but anti-correlates with the disc mass accretion rate. As pointed out earlier, the size of the corona increases or decreases depending on the interplay between the disc and halo mass accretion rates. Therefore, all these changes in corona geometry, spectral state, QPO frequency, and its Q-factor are interlinked through the fundamental flow parameters of accretion, which are the mass accretion rate. 

The source exhibited `flip-flop' behaviour in one of the observations, however, the physical reason is unclear. There are several explanations for this e.g., accretion disc instability, ejection of jets, fluctuation in viscosity etc \citep{Miy91, Tak97, Lyubarskii1997MNRAS.292..679L, Lasota2001NewAR..45..449L, FenderEtal2004MNRAS.355.1105F, Sri12, SritamEtal2016ApJ...823...67S}. We note that, during the {\it AstroSat} observation period, the detection of jet ejection or radio flare was not reported \citep{BogensbergerEtal2020A&A...641A.101B}. However, the \jetcaf\, model fitted parameters do show the signature of outflows, from the observation 2, when the `flip-flop' behaviour was triggered. The higher value of $f_{\rm col}$ may indicate that the outflow was not likely to be a collimated jet, rather wind type. In addition, from the \jetcaf\, model fits, we can see that the accretion rate changed significantly, which is sufficient to change the flux behaviour of the source \textcolor{black}{(see \autoref{fig:SpecNH0}, the dashed and dash-dotted lines for segments 1 and 2)}. Therefore, we argue that the `flip-flop' may originate either due to the change in accretion rates, possibly when the disc accretion rate dominates over the halo accretion rate or due to ejection of mass outflows, which significantly obscured the central continuum emission. Furthermore, all the characteristics mentioned above might be the consequences of the same underlying mechanism, which is the change in the mass accretion rates.

Modelling the broadband X-ray spectra with the \jetcaf\, model helps us estimate the binary parameters for the system. We used the disc mass accretion rate from the \jetcaf\, model and fitted the long-term {\it MAXI} light curve of the source to derive the binary parameters. The decay phase of the light curve is considered for this analysis, and the estimated decay timescale (viscous decay) is $\sim 77$ day for the source. As the viscous timescale depends on the outer radius of the disc, it connects both the orbital separation and the period of the system for a given binary mass ratio. Our estimate of binary separation ($a$) varies between $2.0-2.3\times 10^{11}$ cm, for the companion mass of 0.5 -- 0.8 $M_\odot$, by assuming the companion as a mid or late K-type star \citep{BahramianEtal2018ATel11342....1B}. The orbital period of the companion comes out to be between 3.6 to 4.4 hr. The estimated short orbital period of Swift\,J1658.2--4242 is consistent with other BHXRB systems \citep[3.2--6.3 hr;][and references therein]{ZuritaEtal2008ApJ...681.1458Z, RussellEtal2014MNRAS.439.1381R, JaisawalEtal2015ATel.7361....1J, JesusEtal2016A&A...587A..61C}. Detailed optical photometric studies using large telescopes are required to verify the estimated orbital period of the source.

\section*{Acknowledgements}
We thank the anonymous referee for making insightful comments and suggestions that improved the paper. SM acknowledges Ramanujan Fellowship (\#RJF/2020/000113) by SERB-DST, Govt. of India, for this research. VJ would like to acknowledge the Centre for Research, CHRIST (Deemed to be University), for the financial support in the form of a Seed Money Grant (SMSS-2217).
The research is based on the results obtained from the {\it AstroSat} mission of the Indian Space Research Organization (ISRO), archived at the Indian Space Science Data Centre (ISSDC). This work has used the data from the LAXPC and SXT instruments. We thank the LAXPC Payload Operation Center (POC) and the SXT POC at TIFR, Mumbai, for providing the data via the ISSDC data archive and the necessary software tools.

%%%%%%%%%%%%%%%%%%%%%%%%%%%%%%%%%%%%%%%%%%%%%%%%%%
\section*{Data Availability}

The data used in this article are available in the ISRO's Science Data Archive for {\it AstroSat} Mission (\url{https://astrobrowse.issdc.gov.in/astro_archive/archive/Home.jsp}). The model used in the paper can be shared on reasonable request to the corresponding author.



\bibliographystyle{mnras}
\bibliography{reference} % if your bibtex file is called example.bib

\begin{thebibliography}{}
\makeatletter
\relax
\def\mn@urlcharsother{\let\do\@makeother \do\$\do\&\do\#\do\^\do\_\do\%\do\~}
\def\mn@doi{\begingroup\mn@urlcharsother \@ifnextchar [ {\mn@doi@}
  {\mn@doi@[]}}
\def\mn@doi@[#1]#2{\def\@tempa{#1}\ifx\@tempa\@empty \href
  {http://dx.doi.org/#2} {doi:#2}\else \href {http://dx.doi.org/#2} {#1}\fi
  \endgroup}
\def\mn@eprint#1#2{\mn@eprint@#1:#2::\@nil}
\def\mn@eprint@arXiv#1{\href {http://arxiv.org/abs/#1} {{\tt arXiv:#1}}}
\def\mn@eprint@dblp#1{\href {http://dblp.uni-trier.de/rec/bibtex/#1.xml}
  {dblp:#1}}
\def\mn@eprint@#1:#2:#3:#4\@nil{\def\@tempa {#1}\def\@tempb {#2}\def\@tempc
  {#3}\ifx \@tempc \@empty \let \@tempc \@tempb \let \@tempb \@tempa \fi \ifx
  \@tempb \@empty \def\@tempb {arXiv}\fi \@ifundefined
  {mn@eprint@\@tempb}{\@tempb:\@tempc}{\expandafter \expandafter \csname
  mn@eprint@\@tempb\endcsname \expandafter{\@tempc}}}

\bibitem[\protect\citeauthoryear{{Agrawal} et~al.,}{{Agrawal}
  et~al.}{2017}]{AgrawalEtal2017JApA...38...30A}
{Agrawal} P.~C.,  et~al., 2017, \mn@doi [Journal of Astrophysics and Astronomy]
  {10.1007/s12036-017-9451-z}, \href
  {https://ui.adsabs.harvard.edu/abs/2017JApA...38...30A} {38, 30}

\bibitem[\protect\citeauthoryear{{Antia} et~al.,}{{Antia} et~al.}{2017}]{Ant17}
{Antia} H.~M.,  et~al., 2017, \mn@doi [\apjs] {10.3847/1538-4365/aa7a0e}, \href
  {http://adsabs.harvard.edu/abs/2017ApJS..231...10A} {231, 10}

\bibitem[\protect\citeauthoryear{{Arnaud}}{{Arnaud}}{1996}]{Arnaud1996ASPC..101...17A}
{Arnaud} K.~A.,  1996, in {Jacoby} G.~H.,  {Barnes} J.,  eds,  Astronomical
  Society of the Pacific Conference Series Vol. 101, Astronomical Data Analysis
  Software and Systems V. p.~17

\bibitem[\protect\citeauthoryear{{Baby}, {Bhuvana}, {Radhika}, {Katoch},
  {Mandal}  \& {Nandi}}{{Baby} et~al.}{2021}]{BabyEtal2021MNRAS.508.2447B}
{Baby} B.~E.,  {Bhuvana} G.~R.,  {Radhika} D.,  {Katoch} T.,  {Mandal} S.,
  {Nandi} A.,  2021, \mn@doi [\mnras] {10.1093/mnras/stab2719}, \href
  {https://ui.adsabs.harvard.edu/abs/2021MNRAS.508.2447B} {508, 2447}

\bibitem[\protect\citeauthoryear{{Bahramian}, {Britt}  \&
  {Strader}}{{Bahramian} et~al.}{2018}]{BahramianEtal2018ATel11342....1B}
{Bahramian} A.,  {Britt} C.,   {Strader} J.,  2018, The Astronomer's Telegram,
  \href {https://ui.adsabs.harvard.edu/abs/2018ATel11342....1B} {11342, 1}

\bibitem[\protect\citeauthoryear{{Banerjee}, {Bhattacharjee}, {Chatterjee},
  {Debnath}, {Chakrabarti}, {Katoch}  \& {Antia}}{{Banerjee}
  et~al.}{2021}]{Ban21}
{Banerjee} A.,  {Bhattacharjee} A.,  {Chatterjee} D.,  {Debnath} D.,
  {Chakrabarti} S.~K.,  {Katoch} T.,   {Antia} H.~M.,  2021, \mn@doi [\apj]
  {10.3847/1538-4357/ac0150}, \href
  {https://ui.adsabs.harvard.edu/abs/2021ApJ...916...68B} {916, 68}

\bibitem[\protect\citeauthoryear{{Barthelmy}, {D'Avanzo}, {Deich}, {Gronwall},
  {Melandri}, {Page}  \& {Palmer}}{{Barthelmy} et~al.}{2018}]{Bar18}
{Barthelmy} S.~D.,  {D'Avanzo} P.,  {Deich} A.,  {Gronwall} C.,  {Melandri} A.,
   {Page} K.~L.,   {Palmer} D.~M.,  2018, GRB Coordinates Network, \href
  {https://ui.adsabs.harvard.edu/abs/2018GCN.22416....1B} {22416, 1}

\bibitem[\protect\citeauthoryear{{Belloni}, {Homan}, {Casella}, {van der Klis},
  {Nespoli}, {Lewin}, {Miller}  \& {M{\'e}ndez}}{{Belloni}
  et~al.}{2005}]{BelloniEtal2005A&A...440..207B}
{Belloni} T.,  {Homan} J.,  {Casella} P.,  {van der Klis} M.,  {Nespoli} E.,
  {Lewin} W.~H.~G.,  {Miller} J.~M.,   {M{\'e}ndez} M.,  2005, \mn@doi [\aap]
  {10.1051/0004-6361:20042457}, \href
  {https://ui.adsabs.harvard.edu/abs/2005A&A...440..207B} {440, 207}

\bibitem[\protect\citeauthoryear{{Beri} \& {Altamirano}}{{Beri} \&
  {Altamirano}}{2018}]{BeriAlta2018ATel12072....1B}
{Beri} A.,  {Altamirano} D.,  2018, The Astronomer's Telegram, \href
  {https://ui.adsabs.harvard.edu/abs/2018ATel12072....1B} {12072, 1}

\bibitem[\protect\citeauthoryear{{Bhargava}, {Belloni}, {Bhattacharya}  \&
  {Misra}}{{Bhargava} et~al.}{2019}]{Bha19}
{Bhargava} Y.,  {Belloni} T.,  {Bhattacharya} D.,   {Misra} R.,  2019, \mn@doi
  [\mnras] {10.1093/mnras/stz1774}, \href
  {https://ui.adsabs.harvard.edu/abs/2019MNRAS.488..720B} {488, 720}

\bibitem[\protect\citeauthoryear{{Bogensberger} et~al.,}{{Bogensberger}
  et~al.}{2020}]{BogensbergerEtal2020A&A...641A.101B}
{Bogensberger} D.,  et~al., 2020, \mn@doi [\aap] {10.1051/0004-6361/202037657},
  \href {https://ui.adsabs.harvard.edu/abs/2020A&A...641A.101B} {641, A101}

\bibitem[\protect\citeauthoryear{{Cabanac}, {Henri}, {Petrucci}, {Malzac},
  {Ferreira}  \& {Belloni}}{{Cabanac}
  et~al.}{2010}]{CabanacEtal2010MNRAS.404..738C}
{Cabanac} C.,  {Henri} G.,  {Petrucci} P.~O.,  {Malzac} J.,  {Ferreira} J.,
  {Belloni} T.~M.,  2010, \mn@doi [\mnras] {10.1111/j.1365-2966.2010.16340.x},
  \href {https://ui.adsabs.harvard.edu/abs/2010MNRAS.404..738C} {404, 738}

\bibitem[\protect\citeauthoryear{{Cannizzo}}{{Cannizzo}}{1994}]{Cannizzo1994ApJ...435..389C}
{Cannizzo} J.~K.,  1994, \mn@doi [\apj] {10.1086/174821}, \href
  {https://ui.adsabs.harvard.edu/abs/1994ApJ...435..389C} {435, 389}

\bibitem[\protect\citeauthoryear{{Casella}, {Belloni}  \& {Stella}}{{Casella}
  et~al.}{2005}]{CasellaEtal2005ApJ...629..403C}
{Casella} P.,  {Belloni} T.,   {Stella} L.,  2005, \mn@doi [\apj]
  {10.1086/431174}, \href
  {https://ui.adsabs.harvard.edu/abs/2005ApJ...629..403C} {629, 403}

\bibitem[\protect\citeauthoryear{{Chakrabarti}}{{Chakrabarti}}{1989}]{Chakrabarti1989ApJ...347..365C}
{Chakrabarti} S.~K.,  1989, \mn@doi [\apj] {10.1086/168125}, \href
  {https://ui.adsabs.harvard.edu/abs/1989ApJ...347..365C} {347, 365}

\bibitem[\protect\citeauthoryear{{Chakrabarti}}{{Chakrabarti}}{1999}]{Chakrabarti1999}
{Chakrabarti} S.~K.,  1999, \aap, \href
  {https://ui.adsabs.harvard.edu/abs/1999A&A...351..185C} {351, 185}

\bibitem[\protect\citeauthoryear{{Chakrabarti} \& {Molteni}}{{Chakrabarti} \&
  {Molteni}}{1993}]{ChakrabartiMolteni1993ApJ...417..671C}
{Chakrabarti} S.~K.,  {Molteni} D.,  1993, \mn@doi [\apj] {10.1086/173345},
  \href {https://ui.adsabs.harvard.edu/abs/1993ApJ...417..671C} {417, 671}

\bibitem[\protect\citeauthoryear{{Chakrabarti} \& {Titarchuk}}{{Chakrabarti} \&
  {Titarchuk}}{1995}]{ChakTitarchuk1995ApJ...455..623C}
{Chakrabarti} S.,  {Titarchuk} L.~G.,  1995, \mn@doi [\apj] {10.1086/176610},
  \href {https://ui.adsabs.harvard.edu/abs/1995ApJ...455..623C} {455, 623}

\bibitem[\protect\citeauthoryear{{Chakrabarti} \& {Wiita}}{{Chakrabarti} \&
  {Wiita}}{1993}]{ChakrabartiWiita1993A&A...271..216C}
{Chakrabarti} S.~K.,  {Wiita} P.~J.,  1993, \aap, \href
  {https://ui.adsabs.harvard.edu/abs/1993A&A...271..216C} {271, 216}

\bibitem[\protect\citeauthoryear{{Chakrabarti}, {Debnath}, {Nandi}  \&
  {Pal}}{{Chakrabarti} et~al.}{2008}]{chakrabartiEtal2008A&A...489L..41C}
{Chakrabarti} S.~K.,  {Debnath} D.,  {Nandi} A.,   {Pal} P.~S.,  2008, \mn@doi
  [A\&A] {10.1051/0004-6361:200810136}, \href
  {https://ui.adsabs.harvard.edu/abs/2008A&A...489L..41C} {489, L41}

\bibitem[\protect\citeauthoryear{{Chakrabarti}, {Mondal}  \&
  {Debnath}}{{Chakrabarti} et~al.}{2015}]{ChakrabartiEtal2015MNRAS.452.3451C}
{Chakrabarti} S.~K.,  {Mondal} S.,   {Debnath} D.,  2015, \mn@doi [\mnras]
  {10.1093/mnras/stv1566}, \href
  {https://ui.adsabs.harvard.edu/abs/2015MNRAS.452.3451C} {452, 3451}

\bibitem[\protect\citeauthoryear{{Chakrabarti}, {Debnath}  \&
  {Nagarkoti}}{{Chakrabarti} et~al.}{2019}]{ChakrabartiEtal2019AdSpR..63.3749C}
{Chakrabarti} S.~K.,  {Debnath} D.,   {Nagarkoti} S.,  2019, \mn@doi [Advances
  in Space Research] {10.1016/j.asr.2019.02.014}, \href
  {https://ui.adsabs.harvard.edu/abs/2019AdSpR..63.3749C} {63, 3749}

\bibitem[\protect\citeauthoryear{{Chakraborty}, {Navale}, {Ratheesh}  \&
  {Bhattacharyya}}{{Chakraborty} et~al.}{2020}]{Cha20}
{Chakraborty} S.,  {Navale} N.,  {Ratheesh} A.,   {Bhattacharyya} S.,  2020,
  \mn@doi [\mnras] {10.1093/mnras/staa2711}, \href
  {https://ui.adsabs.harvard.edu/abs/2020MNRAS.498.5873C} {498, 5873}

\bibitem[\protect\citeauthoryear{{Chatterjee}, {Debnath}, {Chakrabarti},
  {Mondal}  \& {Jana}}{{Chatterjee}
  et~al.}{2016}]{ChatterjeeEtal2016ApJ...827...88C}
{Chatterjee} D.,  {Debnath} D.,  {Chakrabarti} S.~K.,  {Mondal} S.,   {Jana}
  A.,  2016, \mn@doi [\apj] {10.3847/0004-637X/827/1/88}, \href
  {https://ui.adsabs.harvard.edu/abs/2016ApJ...827...88C} {827, 88}

\bibitem[\protect\citeauthoryear{{Corral-Santana}, {Casares},
  {Mu{\~n}oz-Darias}, {Bauer}, {Mart{\'\i}nez-Pais}  \&
  {Russell}}{{Corral-Santana} et~al.}{2016}]{JesusEtal2016A&A...587A..61C}
{Corral-Santana} J.~M.,  {Casares} J.,  {Mu{\~n}oz-Darias} T.,  {Bauer} F.~E.,
  {Mart{\'\i}nez-Pais} I.~G.,   {Russell} D.~M.,  2016, \mn@doi [\aap]
  {10.1051/0004-6361/201527130}, \href
  {https://ui.adsabs.harvard.edu/abs/2016A&A...587A..61C} {587, A61}

\bibitem[\protect\citeauthoryear{{Debnath}, {Chakrabarti}  \&
  {Mondal}}{{Debnath} et~al.}{2014}]{Debnathetal2014}
{Debnath} D.,  {Chakrabarti} S.~K.,   {Mondal} S.,  2014, \mn@doi [\mnras]
  {10.1093/mnrasl/slu024}, \href
  {https://ui.adsabs.harvard.edu/abs/2014MNRAS.440L.121D} {440, L121}

\bibitem[\protect\citeauthoryear{{Debnath}, {Mondal}  \&
  {Chakrabarti}}{{Debnath} et~al.}{2015}]{DebnathElal2015MNRAS.447.1984D}
{Debnath} D.,  {Mondal} S.,   {Chakrabarti} S.~K.,  2015, \mn@doi [\mnras]
  {10.1093/mnras/stu2588}, \href
  {https://ui.adsabs.harvard.edu/abs/2015MNRAS.447.1984D} {447, 1984}

\bibitem[\protect\citeauthoryear{{Done}, {Tomaru}  \& {Takahashi}}{{Done}
  et~al.}{2018}]{DoneEtal2018MNRAS.473..838D}
{Done} C.,  {Tomaru} R.,   {Takahashi} T.,  2018, \mn@doi [\mnras]
  {10.1093/mnras/stx2400}, \href
  {https://ui.adsabs.harvard.edu/abs/2018MNRAS.473..838D} {473, 838}

\bibitem[\protect\citeauthoryear{{Dubus}, {Lasota}, {Hameury}  \&
  {Charles}}{{Dubus} et~al.}{1999}]{DubusEtal1999MNRAS.303..139D}
{Dubus} G.,  {Lasota} J.-P.,  {Hameury} J.-M.,   {Charles} P.,  1999, \mn@doi
  [\mnras] {10.1046/j.1365-8711.1999.02212.x}, \href
  {https://ui.adsabs.harvard.edu/abs/1999MNRAS.303..139D} {303, 139}

\bibitem[\protect\citeauthoryear{{Eggleton}}{{Eggleton}}{1983}]{Eggleton1983ApJ...268..368E}
{Eggleton} P.~P.,  1983, \mn@doi [\apj] {10.1086/160960}, \href
  {https://ui.adsabs.harvard.edu/abs/1983ApJ...268..368E} {268, 368}

\bibitem[\protect\citeauthoryear{{Fender}, {Belloni}  \& {Gallo}}{{Fender}
  et~al.}{2004}]{FenderEtal2004MNRAS.355.1105F}
{Fender} R.~P.,  {Belloni} T.~M.,   {Gallo} E.,  2004, \mn@doi [\mnras]
  {10.1111/j.1365-2966.2004.08384.x}, \href
  {https://ui.adsabs.harvard.edu/abs/2004MNRAS.355.1105F} {355, 1105}

\bibitem[\protect\citeauthoryear{Frank, King  \& Raine}{Frank
  et~al.}{2002}]{FrankBook2002}
Frank J.,  King A.,   Raine D.,  2002, Accretion Power in Astrophysics, 3 edn.
Cambridge University Press, \mn@doi{10.1017/CBO9781139164245}

\bibitem[\protect\citeauthoryear{{Fukue}}{{Fukue}}{1987}]{Fukue1987PASJ...39..309F}
{Fukue} J.,  1987, \pasj, \href
  {https://ui.adsabs.harvard.edu/abs/1987PASJ...39..309F} {39, 309}

\bibitem[\protect\citeauthoryear{{Galeev}, {Rosner}  \& {Vaiana}}{{Galeev}
  et~al.}{1979}]{Galeev1979ApJ...229..318G}
{Galeev} A.~A.,  {Rosner} R.,   {Vaiana} G.~S.,  1979, \mn@doi [\apj]
  {10.1086/156957}, \href
  {https://ui.adsabs.harvard.edu/abs/1979ApJ...229..318G} {229, 318}

\bibitem[\protect\citeauthoryear{{Garain} \& {Kim}}{{Garain} \&
  {Kim}}{2022}]{GarainKim2022arXiv221208310G}
{Garain} S.~K.,  {Kim} J.,  2022, arXiv e-prints, \href
  {https://ui.adsabs.harvard.edu/abs/2022arXiv221208310G} {p. arXiv:2212.08310}

\bibitem[\protect\citeauthoryear{{Garain}, {Ghosh}  \& {Chakrabarti}}{{Garain}
  et~al.}{2014}]{GarainEtal2014MNRAS.437.1329G}
{Garain} S.~K.,  {Ghosh} H.,   {Chakrabarti} S.~K.,  2014, \mn@doi [\mnras]
  {10.1093/mnras/stt1969}, \href
  {https://ui.adsabs.harvard.edu/abs/2014MNRAS.437.1329G} {437, 1329}

\bibitem[\protect\citeauthoryear{{Gilfanov}, {Revnivtsev}  \&
  {Molkov}}{{Gilfanov} et~al.}{2003}]{Gilfanov2003A&A...410..217G}
{Gilfanov} M.,  {Revnivtsev} M.,   {Molkov} S.,  2003, \mn@doi [\aap]
  {10.1051/0004-6361:20031141}, \href
  {https://ui.adsabs.harvard.edu/abs/2003A&A...410..217G} {410, 217}

\bibitem[\protect\citeauthoryear{{Giri} \& {Chakrabarti}}{{Giri} \&
  {Chakrabarti}}{2013}]{GiriChak2013MNRAS.430.2836G}
{Giri} K.,  {Chakrabarti} S.~K.,  2013, \mn@doi [\mnras]
  {10.1093/mnras/stt087}, \href
  {https://ui.adsabs.harvard.edu/abs/2013MNRAS.430.2836G} {430, 2836}

\bibitem[\protect\citeauthoryear{{Grebenev}, {Mereminskiy}, {Prosvetov},
  {Ducci}, {Bozzo}, {Savchenko}  \& {Ferrigno}}{{Grebenev}
  et~al.}{2018}]{GrebenevEtal2018ATel11306....1G}
{Grebenev} S.~A.,  {Mereminskiy} I.~A.,  {Prosvetov} A.~V.,  {Ducci} L.,
  {Bozzo} E.,  {Savchenko} V.,   {Ferrigno} C.,  2018, The Astronomer's
  Telegram, \href {https://ui.adsabs.harvard.edu/abs/2018ATel11306....1G}
  {11306, 1}

\bibitem[\protect\citeauthoryear{{Grinberg}, {Eikmann}, {Kreykenbohm}  \&
  {Wilms}}{{Grinberg} et~al.}{2018}]{GrinbergEtal2018ATel11318....1G}
{Grinberg} V.,  {Eikmann} W.,  {Kreykenbohm} I.,   {Wilms} J.,  2018, The
  Astronomer's Telegram, \href
  {https://ui.adsabs.harvard.edu/abs/2018ATel11318....1G} {11318, 1}

\bibitem[\protect\citeauthoryear{{Homan}, {Wijnands}, {van der Klis},
  {Belloni}, {van Paradijs}, {Klein-Wolt}, {Fender}  \& {M{\'e}ndez}}{{Homan}
  et~al.}{2001}]{HomanEtal2001ApJS..132..377H}
{Homan} J.,  {Wijnands} R.,  {van der Klis} M.,  {Belloni} T.,  {van Paradijs}
  J.,  {Klein-Wolt} M.,  {Fender} R.,   {M{\'e}ndez} M.,  2001, \mn@doi [\apjs]
  {10.1086/318954}, \href
  {https://ui.adsabs.harvard.edu/abs/2001ApJS..132..377H} {132, 377}

\bibitem[\protect\citeauthoryear{{Ingram}, {Done}  \& {Fragile}}{{Ingram}
  et~al.}{2009}]{IngramEtal2009MNRAS.397L.101I}
{Ingram} A.,  {Done} C.,   {Fragile} P.~C.,  2009, \mn@doi [\mnras]
  {10.1111/j.1745-3933.2009.00693.x}, \href
  {https://ui.adsabs.harvard.edu/abs/2009MNRAS.397L.101I} {397, L101}

\bibitem[\protect\citeauthoryear{{Jaisawal}, {Homan}, {Naik}  \&
  {Jonker}}{{Jaisawal} et~al.}{2015}]{JaisawalEtal2015ATel.7361....1J}
{Jaisawal} G.~K.,  {Homan} J.,  {Naik} S.,   {Jonker} P.,  2015, The
  Astronomer's Telegram, \href
  {https://ui.adsabs.harvard.edu/abs/2015ATel.7361....1J} {7361, 1}

\bibitem[\protect\citeauthoryear{{Janiuk}, {Czerny}  \&
  {Siemiginowska}}{{Janiuk} et~al.}{2002}]{JaniukEtal2002ApJ...576..908J}
{Janiuk} A.,  {Czerny} B.,   {Siemiginowska} A.,  2002, \mn@doi [\apj]
  {10.1086/341804}, \href
  {https://ui.adsabs.harvard.edu/abs/2002ApJ...576..908J} {576, 908}

\bibitem[\protect\citeauthoryear{{Jithesh}, {Maqbool}, {Misra}, {T}, {Mall}  \&
  {James}}{{Jithesh} et~al.}{2019}]{JitheshEatal2019ApJ...887..101J}
{Jithesh} V.,  {Maqbool} B.,  {Misra} R.,  {T} A.~R.,  {Mall} G.,   {James} M.,
   2019, \mn@doi [\apj] {10.3847/1538-4357/ab4f6a}, \href
  {https://ui.adsabs.harvard.edu/abs/2019ApJ...887..101J} {887, 101}

\bibitem[\protect\citeauthoryear{{Jithesh}, {Misra}, {Maqbool}  \&
  {Mall}}{{Jithesh} et~al.}{2021}]{Jit21}
{Jithesh} V.,  {Misra} R.,  {Maqbool} B.,   {Mall} G.,  2021, \mn@doi [\mnras]
  {10.1093/mnras/stab1307}, \href
  {https://ui.adsabs.harvard.edu/abs/2021MNRAS.505..713J} {505, 713}

\bibitem[\protect\citeauthoryear{{Kalberla}, {Burton}, {Hartmann}, {Arnal},
  {Bajaja}, {Morras}  \& {P{\"o}ppel}}{{Kalberla}
  et~al.}{2005}]{KalberlaEtal2005A&A...440..775K}
{Kalberla} P.~M.~W.,  {Burton} W.~B.,  {Hartmann} D.,  {Arnal} E.~M.,  {Bajaja}
  E.,  {Morras} R.,   {P{\"o}ppel} W.~G.~L.,  2005, \mn@doi [\aap]
  {10.1051/0004-6361:20041864}, \href
  {https://ui.adsabs.harvard.edu/abs/2005A&A...440..775K} {440, 775}

\bibitem[\protect\citeauthoryear{{Kim}, {Garain}, {Balsara}  \&
  {Chakrabarti}}{{Kim} et~al.}{2017}]{KimEtal2017MNRAS.472..542K}
{Kim} J.,  {Garain} S.~K.,  {Balsara} D.~S.,   {Chakrabarti} S.~K.,  2017,
  \mn@doi [\mnras] {10.1093/mnras/stx1986}, \href
  {https://ui.adsabs.harvard.edu/abs/2017MNRAS.472..542K} {472, 542}

\bibitem[\protect\citeauthoryear{{King}}{{King}}{1998}]{King1998MNRAS.296L..45K}
{King} A.~R.,  1998, \mn@doi [\mnras] {10.1046/j.1365-8711.1998.01652.x}, \href
  {https://ui.adsabs.harvard.edu/abs/1998MNRAS.296L..45K} {296, L45}

\bibitem[\protect\citeauthoryear{{King} \& {Ritter}}{{King} \&
  {Ritter}}{1998}]{KingRitter1998MNRAS.293L..42K}
{King} A.~R.,  {Ritter} H.,  1998, \mn@doi [\mnras]
  {10.1046/j.1365-8711.1998.01295.x}, \href
  {https://ui.adsabs.harvard.edu/abs/1998MNRAS.293L..42K} {293, L42}

\bibitem[\protect\citeauthoryear{{King}, {Pringle}  \& {Livio}}{{King}
  et~al.}{2007}]{KingEtal2007MNRAS.376.1740K}
{King} A.~R.,  {Pringle} J.~E.,   {Livio} M.,  2007, \mn@doi [\mnras]
  {10.1111/j.1365-2966.2007.11556.x}, \href
  {https://ui.adsabs.harvard.edu/abs/2007MNRAS.376.1740K} {376, 1740}

\bibitem[\protect\citeauthoryear{{Lasota}}{{Lasota}}{2001}]{Lasota2001NewAR..45..449L}
{Lasota} J.-P.,  2001, \mn@doi [\nar] {10.1016/S1387-6473(01)00112-9}, \href
  {https://ui.adsabs.harvard.edu/abs/2001NewAR..45..449L} {45, 449}

\bibitem[\protect\citeauthoryear{{Laurent} \& {Titarchuk}}{{Laurent} \&
  {Titarchuk}}{2001}]{LaurentTitarchuk2001ApJ...562L..67L}
{Laurent} P.,  {Titarchuk} L.,  2001, \mn@doi [\apjl] {10.1086/338049}, \href
  {https://ui.adsabs.harvard.edu/abs/2001ApJ...562L..67L} {562, L67}

\bibitem[\protect\citeauthoryear{{Lien} et~al.,}{{Lien}
  et~al.}{2018}]{LienEtal2018GCN.22419....1L}
{Lien} A.~Y.,  et~al., 2018, GRB Coordinates Network, \href
  {https://ui.adsabs.harvard.edu/abs/2018GCN.22419....1L} {22419, 1}

\bibitem[\protect\citeauthoryear{{Lipunova}}{{Lipunova}}{1999}]{Lipunova1999AstL...25..508L}
{Lipunova} G.~V.,  1999, \mn@doi [Astronomy Letters]
  {10.48550/arXiv.astro-ph/9906324}, \href
  {https://ui.adsabs.harvard.edu/abs/1999AstL...25..508L} {25, 508}

\bibitem[\protect\citeauthoryear{{Lyubarskii}}{{Lyubarskii}}{1997}]{Lyubarskii1997MNRAS.292..679L}
{Lyubarskii} Y.~E.,  1997, \mn@doi [\mnras] {10.1093/mnras/292.3.679}, \href
  {https://ui.adsabs.harvard.edu/abs/1997MNRAS.292..679L} {292, 679}

\bibitem[\protect\citeauthoryear{{Majumder}, {Sreehari}, {Aftab}, {Katoch},
  {Das}  \& {Nandi}}{{Majumder} et~al.}{2022}]{Maj22}
{Majumder} S.,  {Sreehari} H.,  {Aftab} N.,  {Katoch} T.,  {Das} S.,   {Nandi}
  A.,  2022, \mn@doi [\mnras] {10.1093/mnras/stac615}, \href
  {https://ui.adsabs.harvard.edu/abs/2022MNRAS.512.2508M} {512, 2508}

\bibitem[\protect\citeauthoryear{{Mall}, {Vadakkumthani}  \& {Misra}}{{Mall}
  et~al.}{2023}]{Mall23}
{Mall} G.,  {Vadakkumthani} J.,   {Misra} R.,  2023, \mn@doi [Research in
  Astronomy and Astrophysics] {10.1088/1674-4527/aca505}, \href
  {https://ui.adsabs.harvard.edu/abs/2023RAA....23a5015M} {23, 015015}

\bibitem[\protect\citeauthoryear{{Maqbool} et~al.,}{{Maqbool}
  et~al.}{2019}]{Maq19}
{Maqbool} B.,  et~al., 2019, \mn@doi [\mnras] {10.1093/mnras/stz930}, \href
  {https://ui.adsabs.harvard.edu/abs/2019MNRAS.486.2964M} {486, 2964}

\bibitem[\protect\citeauthoryear{{Meyer-Hofmeister}, {Liu}  \&
  {Meyer}}{{Meyer-Hofmeister}
  et~al.}{2005}]{Meyer-HofmeisterEtal2005A&A...432..181M}
{Meyer-Hofmeister} E.,  {Liu} B.~F.,   {Meyer} F.,  2005, \mn@doi [\aap]
  {10.1051/0004-6361:20041631}, \href
  {https://ui.adsabs.harvard.edu/abs/2005A&A...432..181M} {432, 181}

\bibitem[\protect\citeauthoryear{{Misra} et~al.,}{{Misra} et~al.}{2017}]{Mis17}
{Misra} R.,  et~al., 2017, \mn@doi [\apj] {10.3847/1538-4357/835/2/195}, \href
  {https://ui.adsabs.harvard.edu/abs/2017ApJ...835..195M} {835, 195}

\bibitem[\protect\citeauthoryear{{Miyamoto}, {Kimura}, {Kitamoto}, {Dotani}  \&
  {Ebisawa}}{{Miyamoto} et~al.}{1991}]{Miy91}
{Miyamoto} S.,  {Kimura} K.,  {Kitamoto} S.,  {Dotani} T.,   {Ebisawa} K.,
  1991, \mn@doi [\apj] {10.1086/170837}, \href
  {https://ui.adsabs.harvard.edu/abs/1991ApJ...383..784M} {383, 784}

\bibitem[\protect\citeauthoryear{{Molla}, {Debnath}, {Chakrabarti}, {Mondal}
  \& {Jana}}{{Molla} et~al.}{2016}]{MollaEtal2016MNRAS.460.3163M}
{Molla} A.~A.,  {Debnath} D.,  {Chakrabarti} S.~K.,  {Mondal} S.,   {Jana} A.,
  2016, \mn@doi [\mnras] {10.1093/mnras/stw860}, \href
  {https://ui.adsabs.harvard.edu/abs/2016MNRAS.460.3163M} {460, 3163}

\bibitem[\protect\citeauthoryear{{Molla}, {Chakrabarti}, {Debnath}  \&
  {Mondal}}{{Molla} et~al.}{2017}]{MollaEtal2017ApJ...834...88M}
{Molla} A.~A.,  {Chakrabarti} S.~K.,  {Debnath} D.,   {Mondal} S.,  2017,
  \mn@doi [\apj] {10.3847/1538-4357/834/1/88}, \href
  {https://ui.adsabs.harvard.edu/abs/2017ApJ...834...88M} {834, 88}

\bibitem[\protect\citeauthoryear{{Molteni}, {Sponholz}  \&
  {Chakrabarti}}{{Molteni} et~al.}{1996}]{MolteniEtal1996ApJ...457..805M}
{Molteni} D.,  {Sponholz} H.,   {Chakrabarti} S.~K.,  1996, \mn@doi [\apj]
  {10.1086/176775}, \href
  {https://ui.adsabs.harvard.edu/abs/1996ApJ...457..805M} {457, 805}

\bibitem[\protect\citeauthoryear{{Mondal}}{{Mondal}}{2020}]{Mondal2020AdSpR..65..693M}
{Mondal} S.,  2020, \mn@doi [Advances in Space Research]
  {10.1016/j.asr.2019.08.002}, \href
  {https://ui.adsabs.harvard.edu/abs/2020AdSpR..65..693M} {65, 693}

\bibitem[\protect\citeauthoryear{{Mondal} \& {Chakrabarti}}{{Mondal} \&
  {Chakrabarti}}{2021}]{MondalChak2021ApJ...920...41M}
{Mondal} S.,  {Chakrabarti} S.~K.,  2021, \mn@doi [\apj]
  {10.3847/1538-4357/ac14c2}, \href
  {https://ui.adsabs.harvard.edu/abs/2021ApJ...920...41M} {920, 41}

\bibitem[\protect\citeauthoryear{{Mondal}, {Debnath}  \&
  {Chakrabarti}}{{Mondal} et~al.}{2014}]{MondalEtal2014ApJ...786....4M}
{Mondal} S.,  {Debnath} D.,   {Chakrabarti} S.~K.,  2014, \mn@doi [\apj]
  {10.1088/0004-637X/786/1/4}, \href
  {https://ui.adsabs.harvard.edu/abs/2014ApJ...786....4M} {786, 4}

\bibitem[\protect\citeauthoryear{{Mondal}, {Chakrabarti}, {Nagarkoti}  \&
  {Ar{\'e}valo}}{{Mondal} et~al.}{2017}]{MondalEtal2017ApJ...850...47M}
{Mondal} S.,  {Chakrabarti} S.~K.,  {Nagarkoti} S.,   {Ar{\'e}valo} P.,  2017,
  \mn@doi [\apj] {10.3847/1538-4357/aa7e27}, \href
  {https://ui.adsabs.harvard.edu/abs/2017ApJ...850...47M} {850, 47}

\bibitem[\protect\citeauthoryear{{Motta}, {Mu{\~n}oz-Darias}, {Casella},
  {Belloni}  \& {Homan}}{{Motta} et~al.}{2011}]{MottaEtal2011MNRAS.418.2292M}
{Motta} S.,  {Mu{\~n}oz-Darias} T.,  {Casella} P.,  {Belloni} T.,   {Homan} J.,
   2011, \mn@doi [\mnras] {10.1111/j.1365-2966.2011.19566.x}, \href
  {https://ui.adsabs.harvard.edu/abs/2011MNRAS.418.2292M} {418, 2292}

\bibitem[\protect\citeauthoryear{{Nandi}, {Debnath}, {Mandal}  \&
  {Chakrabarti}}{{Nandi} et~al.}{2012}]{NandiEtal2012A&A...542A..56N}
{Nandi} A.,  {Debnath} D.,  {Mandal} S.,   {Chakrabarti} S.~K.,  2012, \mn@doi
  [\aap] {10.1051/0004-6361/201117844}, \href
  {https://ui.adsabs.harvard.edu/abs/2012A&A...542A..56N} {542, A56}

\bibitem[\protect\citeauthoryear{{Narayan}, {McClintock}  \& {Yi}}{{Narayan}
  et~al.}{1996}]{NarayanEtal1996ApJ...457..821N}
{Narayan} R.,  {McClintock} J.~E.,   {Yi} I.,  1996, \mn@doi [\apj]
  {10.1086/176777}, \href
  {https://ui.adsabs.harvard.edu/abs/1996ApJ...457..821N} {457, 821}

\bibitem[\protect\citeauthoryear{{Reeves}, {Done}, {Pounds}, {Terashima},
  {Hayashida}, {Anabuki}, {Uchino}  \& {Turner}}{{Reeves}
  et~al.}{2008}]{ReevesEtal2008MNRAS.385L.108R}
{Reeves} J.,  {Done} C.,  {Pounds} K.,  {Terashima} Y.,  {Hayashida} K.,
  {Anabuki} N.,  {Uchino} M.,   {Turner} M.,  2008, \mn@doi [\mnras]
  {10.1111/j.1745-3933.2008.00443.x}, \href
  {https://ui.adsabs.harvard.edu/abs/2008MNRAS.385L.108R} {385, L108}

\bibitem[\protect\citeauthoryear{{Remillard} \& {McClintock}}{{Remillard} \&
  {McClintock}}{2006}]{RemillardMcClin2006ARA&A..44...49R}
{Remillard} R.~A.,  {McClintock} J.~E.,  2006, \mn@doi [\araa]
  {10.1146/annurev.astro.44.051905.092532}, \href
  {https://ui.adsabs.harvard.edu/abs/2006ARA&A..44...49R} {44, 49}

\bibitem[\protect\citeauthoryear{{Russell}, {Soria}, {Motch}, {Pakull},
  {Torres}, {Curran}, {Jonker}  \& {Miller-Jones}}{{Russell}
  et~al.}{2014}]{RussellEtal2014MNRAS.439.1381R}
{Russell} T.~D.,  {Soria} R.,  {Motch} C.,  {Pakull} M.~W.,  {Torres} M.~A.~P.,
   {Curran} P.~A.,  {Jonker} P.~G.,   {Miller-Jones} J.~C.~A.,  2014, \mn@doi
  [\mnras] {10.1093/mnras/stt2480}, \href
  {https://ui.adsabs.harvard.edu/abs/2014MNRAS.439.1381R} {439, 1381}

\bibitem[\protect\citeauthoryear{{Russell}, {Miller-Jones}, {Sivakoff}  \&
  {Tetarenko}}{{Russell} et~al.}{2018}]{RussellEtal2018ATel11322....1R}
{Russell} T.~D.,  {Miller-Jones} J.~C.~A.,  {Sivakoff} G.~R.,   {Tetarenko}
  A.~J.,  2018, The Astronomer's Telegram, \href
  {https://ui.adsabs.harvard.edu/abs/2018ATel11322....1R} {11322, 1}

\bibitem[\protect\citeauthoryear{{Ryu}, {Chakrabarti}  \& {Molteni}}{{Ryu}
  et~al.}{1997}]{RyuEtal1997ApJ...474..378R}
{Ryu} D.,  {Chakrabarti} S.~K.,   {Molteni} D.,  1997, \mn@doi [\apj]
  {10.1086/303461}, \href
  {https://ui.adsabs.harvard.edu/abs/1997ApJ...474..378R} {474, 378}

\bibitem[\protect\citeauthoryear{{Shakura} \& {Sunyaev}}{{Shakura} \&
  {Sunyaev}}{1973}]{ShakuraSunyaev1973A&A....24..337S}
{Shakura} N.~I.,  {Sunyaev} R.~A.,  1973, \aap, \href
  {https://ui.adsabs.harvard.edu/abs/1973A&A....24..337S} {24, 337}

\bibitem[\protect\citeauthoryear{{Shui} et~al.,}{{Shui}
  et~al.}{2021}]{ShuiEtal2021MNRAS.508..287S}
{Shui} Q.~C.,  et~al., 2021, \mn@doi [\mnras] {10.1093/mnras/stab2521}, \href
  {https://ui.adsabs.harvard.edu/abs/2021MNRAS.508..287S} {508, 287}

\bibitem[\protect\citeauthoryear{{Singh} et~al.,}{{Singh}
  et~al.}{2014}]{SinghEtal2014SPIE.9144E..1SS}
{Singh} K.~P.,  et~al., 2014, in {Takahashi} T.,  {den Herder} J.-W.~A.,
  {Bautz} M.,  eds,  Society of Photo-Optical Instrumentation Engineers (SPIE)
  Conference Series Vol. 9144, Space Telescopes and Instrumentation 2014:
  Ultraviolet to Gamma Ray. p. 91441S, \mn@doi{10.1117/12.2062667}

\bibitem[\protect\citeauthoryear{{Singh} et~al.,}{{Singh} et~al.}{2016}]{Sin16}
{Singh} K.~P.,  et~al., 2016, in Space Telescopes and Instrumentation 2016:
  Ultraviolet to Gamma Ray. p. 99051E, \mn@doi{10.1117/12.2235309}

\bibitem[\protect\citeauthoryear{{Singh} et~al.,}{{Singh} et~al.}{2017}]{Sin17}
{Singh} K.~P.,  et~al., 2017, \mn@doi [Journal of Astrophysics and Astronomy]
  {10.1007/s12036-017-9448-7}, \href
  {http://adsabs.harvard.edu/abs/2017JApA...38...29S} {38, 29}

\bibitem[\protect\citeauthoryear{{Sreehari}, {Nandi}, {Das}, {Agrawal},
  {Mandal}, {Ramadevi}  \& {Katoch}}{{Sreehari} et~al.}{2020}]{Sre20}
{Sreehari} H.,  {Nandi} A.,  {Das} S.,  {Agrawal} V.~K.,  {Mandal} S.,
  {Ramadevi} M.~C.,   {Katoch} T.,  2020, \mn@doi [\mnras]
  {10.1093/mnras/staa3135}, \href
  {https://ui.adsabs.harvard.edu/abs/2020MNRAS.499.5891S} {499, 5891}

\bibitem[\protect\citeauthoryear{{Sridhar}, {Bhattacharyya}, {Chandra}  \&
  {Antia}}{{Sridhar} et~al.}{2019}]{Sri19}
{Sridhar} N.,  {Bhattacharyya} S.,  {Chandra} S.,   {Antia} H.~M.,  2019,
  \mn@doi [\mnras] {10.1093/mnras/stz1476}, \href
  {https://ui.adsabs.harvard.edu/abs/2019MNRAS.487.4221S} {487, 4221}

\bibitem[\protect\citeauthoryear{{Sriram}, {Rao}  \& {Choi}}{{Sriram}
  et~al.}{2012}]{Sri12}
{Sriram} K.,  {Rao} A.~R.,   {Choi} C.~S.,  2012, \mn@doi [\aap]
  {10.1051/0004-6361/201218799}, \href
  {https://ui.adsabs.harvard.edu/abs/2012A&A...541A...6S} {541, A6}

\bibitem[\protect\citeauthoryear{{Sriram}, {Rao}  \& {Choi}}{{Sriram}
  et~al.}{2016}]{SritamEtal2016ApJ...823...67S}
{Sriram} K.,  {Rao} A.~R.,   {Choi} C.~S.,  2016, \mn@doi [\apj]
  {10.3847/0004-637X/823/1/67}, \href
  {https://ui.adsabs.harvard.edu/abs/2016ApJ...823...67S} {823, 67}

\bibitem[\protect\citeauthoryear{{Stella} \& {Vietri}}{{Stella} \&
  {Vietri}}{1998}]{StellaVietri1998ApJ...492L..59S}
{Stella} L.,  {Vietri} M.,  1998, \mn@doi [\apjl] {10.1086/311075}, \href
  {https://ui.adsabs.harvard.edu/abs/1998ApJ...492L..59S} {492, L59}

\bibitem[\protect\citeauthoryear{{Stella}, {Vietri}  \& {Morsink}}{{Stella}
  et~al.}{1999}]{StellaEtal1999ApJ...524L..63S}
{Stella} L.,  {Vietri} M.,   {Morsink} S.~M.,  1999, \mn@doi [\apjl]
  {10.1086/312291}, \href
  {https://ui.adsabs.harvard.edu/abs/1999ApJ...524L..63S} {524, L63}

\bibitem[\protect\citeauthoryear{{Tagger} \& {Pellat}}{{Tagger} \&
  {Pellat}}{1999}]{TaggerPellat1999A&A...349.1003T}
{Tagger} M.,  {Pellat} R.,  1999, \aap, \href
  {https://ui.adsabs.harvard.edu/abs/1999A&A...349.1003T} {349, 1003}

\bibitem[\protect\citeauthoryear{{Takizawa} et~al.,}{{Takizawa}
  et~al.}{1997}]{Tak97}
{Takizawa} M.,  et~al., 1997, \mn@doi [\apj] {10.1086/304759}, \href
  {https://ui.adsabs.harvard.edu/abs/1997ApJ...489..272T} {489, 272}

\bibitem[\protect\citeauthoryear{{Titarchuk} \& {Fiorito}}{{Titarchuk} \&
  {Fiorito}}{2004}]{TitarchukFiorito2004ApJ...612..988T}
{Titarchuk} L.,  {Fiorito} R.,  2004, \mn@doi [\apj] {10.1086/422573}, \href
  {https://ui.adsabs.harvard.edu/abs/2004ApJ...612..988T} {612, 988}

\bibitem[\protect\citeauthoryear{{Titarchuk} \& {Shrader}}{{Titarchuk} \&
  {Shrader}}{2005}]{TitarchukShrader2005}
{Titarchuk} L.,  {Shrader} C.,  2005, \mn@doi [\apj] {10.1086/424918}, \href
  {https://ui.adsabs.harvard.edu/abs/2005ApJ...623..362T} {623, 362}

\bibitem[\protect\citeauthoryear{{Wijnands}, {Homan}  \& {van der
  Klis}}{{Wijnands} et~al.}{1999}]{WijnandsEtal1999ApJ...526L..33W}
{Wijnands} R.,  {Homan} J.,   {van der Klis} M.,  1999, \mn@doi [\apjl]
  {10.1086/312365}, \href
  {https://ui.adsabs.harvard.edu/abs/1999ApJ...526L..33W} {526, L33}

\bibitem[\protect\citeauthoryear{{Wilms}, {Allen}  \& {McCray}}{{Wilms}
  et~al.}{2000}]{Wilms2000ApJ...542..914W}
{Wilms} J.,  {Allen} A.,   {McCray} R.,  2000, \mn@doi [\apj] {10.1086/317016},
  \href {https://ui.adsabs.harvard.edu/abs/2000ApJ...542..914W} {542, 914}

\bibitem[\protect\citeauthoryear{{Xiao} et~al.,}{{Xiao}
  et~al.}{2019}]{XiaoEtal2019JHEAp..24...30X}
{Xiao} G.~C.,  et~al., 2019, \mn@doi [Journal of High Energy Astrophysics]
  {10.1016/j.jheap.2019.09.005}, \href
  {https://ui.adsabs.harvard.edu/abs/2019JHEAp..24...30X} {24, 30}

\bibitem[\protect\citeauthoryear{{Xu} et~al.,}{{Xu}
  et~al.}{2018}]{XuEtal2018ApJ...865...18X}
{Xu} Y.,  et~al., 2018, \mn@doi [\apj] {10.3847/1538-4357/aada03}, \href
  {https://ui.adsabs.harvard.edu/abs/2018ApJ...865...18X} {865, 18}

\bibitem[\protect\citeauthoryear{{Xu} et~al.,}{{Xu}
  et~al.}{2019}]{XuEtal2019ApJ...879...93X}
{Xu} Y.,  et~al., 2019, \mn@doi [\apj] {10.3847/1538-4357/ab24bf}, \href
  {https://ui.adsabs.harvard.edu/abs/2019ApJ...879...93X} {879, 93}

\bibitem[\protect\citeauthoryear{{Yadav} et~al.,}{{Yadav}
  et~al.}{2016a}]{Yad16a}
{Yadav} J.~S.,  et~al., 2016a, \mn@doi [\apj] {10.3847/0004-637X/833/1/27},
  \href {http://adsabs.harvard.edu/abs/2016ApJ...833...27Y} {833, 27}

\bibitem[\protect\citeauthoryear{{Yadav} et~al.,}{{Yadav}
  et~al.}{2016b}]{Yad16}
{Yadav} J.~S.,  et~al., 2016b, in Space Telescopes and Instrumentation 2016:
  Ultraviolet to Gamma Ray. p. 99051D, \mn@doi{10.1117/12.2231857}

\bibitem[\protect\citeauthoryear{{Zdziarski}, {Lubi{\'n}ski}, {Gilfanov}  \&
  {Revnivtsev}}{{Zdziarski} et~al.}{2003}]{Zdziarski2003MNRAS.342..355Z}
{Zdziarski} A.~A.,  {Lubi{\'n}ski} P.,  {Gilfanov} M.,   {Revnivtsev} M.,
  2003, \mn@doi [\mnras] {10.1046/j.1365-8711.2003.06556.x}, \href
  {https://ui.adsabs.harvard.edu/abs/2003MNRAS.342..355Z} {342, 355}

\bibitem[\protect\citeauthoryear{{Zurita}, {Durant}, {Torres}, {Shahbaz},
  {Casares}  \& {Steeghs}}{{Zurita}
  et~al.}{2008}]{ZuritaEtal2008ApJ...681.1458Z}
{Zurita} C.,  {Durant} M.,  {Torres} M.~A.~P.,  {Shahbaz} T.,  {Casares} J.,
  {Steeghs} D.,  2008, \mn@doi [\apj] {10.1086/588721}, \href
  {https://ui.adsabs.harvard.edu/abs/2008ApJ...681.1458Z} {681, 1458}

\makeatother
\end{thebibliography}
%\printbibliography
%\begin{thebibliography}{99}
%\end{thebibliography}

%%%%%%%%%%%%%%%%%%%%%%%%%%%%%%%%%%%%%%%%%%%%%%%%%%

%%%%%%%%%%%%%%%%% APPENDICES %%%%%%%%%%%%%%%%%%%%%

\appendix

\section{Best-fit model spectra without absorption} \label{apen:A}
\begin{figure}
%\hspace{-0.5cm}
\begin{tikzpicture}
\draw (0, 0) node[inner sep=0] {\raisebox{0.0cm}{\includegraphics[height=6.5truecm,trim={0.6cm 0.0cm 1.0cm 1.8cm}, clip]{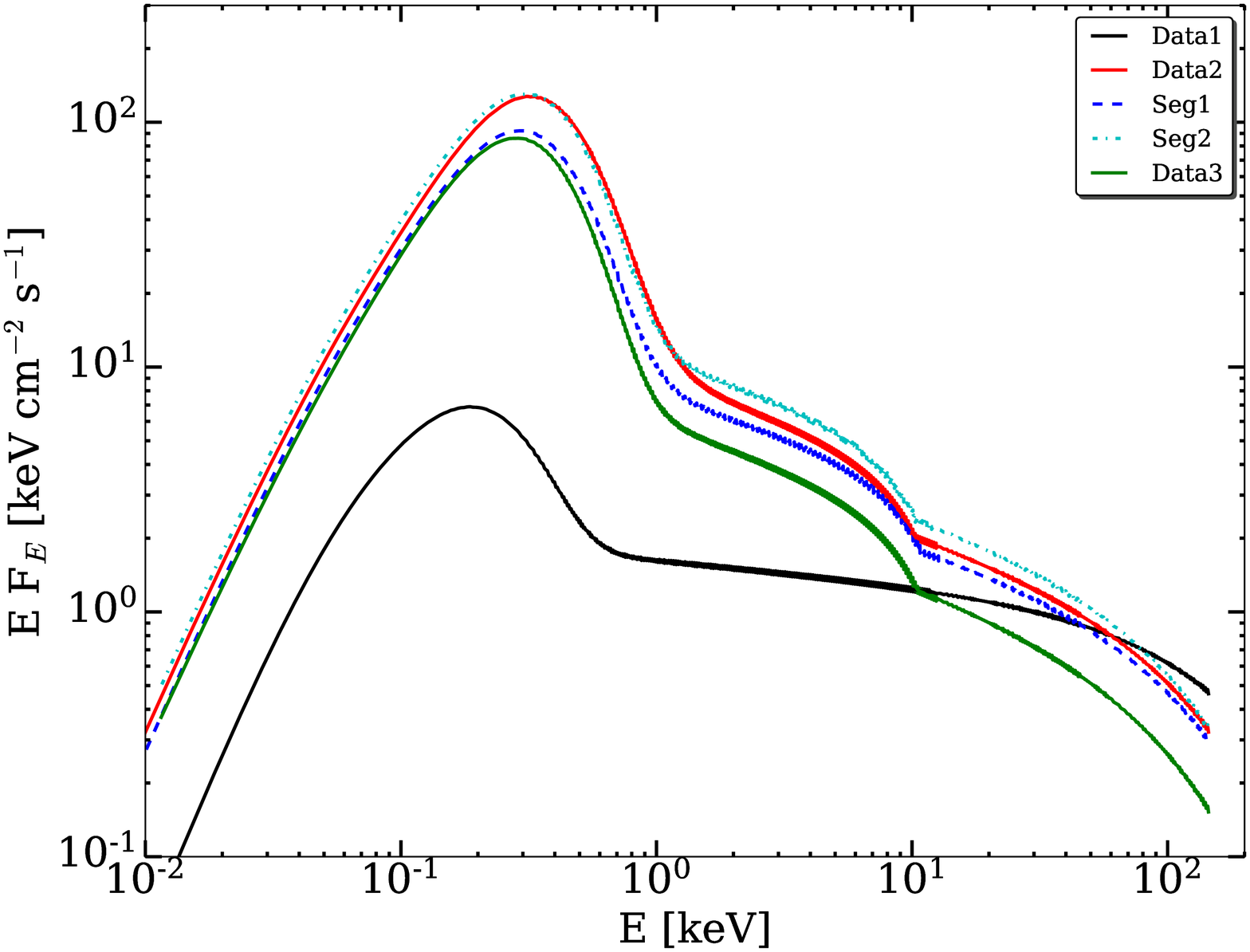}}};
\end{tikzpicture}
\caption{\textcolor{black}{The best-fitted M2 model spectra are shown for all observations and segments. During extracting the model, we set all N$_H$ values to zero and the ``constant" scale-factor for the second spectrum (LAXPC) in XSPEC is set to 1 to keep the model continuous.}}
\label{fig:SpecNH0}
\end{figure}

\textcolor{black}{We have computed the best-fit M2 model spectra for all observations and segments (bottom panel of \autoref{table:jetcafResults}) to see the shape of the continuum spectra without any absorption. During extracting the model spectra, we set all N$_H$, covering fraction, and ionization parameter values to zero and the ``constant" scale-factor for the second spectrum (which is LAXPC in our case) is set to 1 to bring the both model components in the same line. All spectra are shown in \autoref{fig:SpecNH0}, where different colours correspond to different observations and segments. From the figure, it is visible that the significant fraction  of the bolometric flux comes from the {\it AstroSat} energy band.}

%We do not see any excess above 10 keV, which might be an indication of the absence of jets/mass outflows. {\bf do u expect a bump above 10 keV for jet? But we see some residual above 8 keV. Is it contradicting? yes, we do expect some excess in theoretical spectra due to jet. but here we do not see that. The residual might be due to some reflection is there as explained in the main text. BUT REFEREE WANT TO SEE WHERE most of the bolometric emission comes from}} 

% \begin{figure}
% \hspace{-0.5cm}
% \begin{tikzpicture}
% \draw (0, 0) node[inner sep=0] {\raisebox{0.0cm}{\includegraphics[height=6.8truecm,trim={0.6cm 0.0cm 1.0cm 1.8cm}, clip]{Figures/best_fit_mo_plot.eps}}};unabsorbed
% \end{tikzpicture}
% \caption{\textcolor{magenta}{The best-fitted M3 model is shown for all observations and segments.}}
% \label{fig:SpecNH0}
% \end{figure}

%\section{Spectral fit with fixed $N_{HG}$ and no \gabs} \label{apen:B}

%%%%%%%%%%%%%%%%%%%%%%%%%%%%%%%%%%%%%%%%%%%%%%%%%%

% Don't change these lines
\bsp	% typesetting comment
\label{lastpage}
\end{document}